\documentclass[pra,aps,twocolumn,preprintnumbers,amsmath,amssymb, superscriptaddress, showkeys,nofootinbib]{revtex4-1}
\usepackage{amsmath,amsfonts,amstext,txfonts,amssymb,graphics,rotating,
dcolumn,graphics,array,multirow,mathrsfs,gensymb,graphicx,units,braket,xfrac}
\usepackage[T1]{fontenc}
\usepackage{bbold}

\usepackage{natbib}
\usepackage{graphicx} 
\usepackage{epsfig}
\usepackage{epstopdf}
\usepackage{float}
\usepackage{amsmath}
\usepackage{amssymb}
\usepackage{rotating}
\usepackage{lipsum}
\usepackage{xcolor}
\usepackage{mdframed}
 \usepackage{helvet}

\newcommand{\bracket}[2]{\langle #1|#2\rangle}

\def\bea{\begin{eqnarray}}
\def\eea{\end{eqnarray}}
\definecolor{gainsboro}{rgb}{0.86, 0.86, 0.86}

\begin{document}
\title{On the theory of excitonic delocalization for robust vibronic dynamics in LH2}
\author{Felipe Caycedo-Soler}
\affiliation{Institute of Theoretical Physics and Integrated Quantum Science and Technology IQST, University of Ulm, Albert-Einstein-Allee 11, D - 89069 Ulm, Germany}
\email{felipe.caycedo@uni-ulm.de, martin.plenio@uni-ulm.de}
\author{James Lim }
\affiliation{Institute of Theoretical Physics and Integrated Quantum Science and Technology IQST, University of Ulm, Albert-Einstein-Allee 11, D - 89069 Ulm, Germany}
\author{Santiago Oviedo-Casado}
\affiliation{Departmento de F\'isica Aplicada,   Universidad Polit\'ecnica de Cartagena, 30202 Cartagena, Spain}
\author{Niek F. van Hulst}
\affiliation{ICFO - Institut de Ciencies Fotoniques, The Barcelona Institute of Science and Technology, 08860 Castelldefels (Barcelona), Spain and ICREA - Instituci\'o Catalana de Recerca i Estudis Avan\c{c}ats, 08010 Barcelona, Spain}
\author{Susana F. Huelga}
\affiliation{Institute of Theoretical Physics and Integrated Quantum Science and Technology IQST, University of Ulm, Albert-Einstein-Allee 11, D - 89069 Ulm, Germany}
\author{Martin B. Plenio}
\affiliation{Institute of Theoretical Physics and Integrated Quantum Science and Technology IQST, University of Ulm, Albert-Einstein-Allee 11, D - 89069 Ulm, Germany}

\begin{abstract}
Nonlinear spectroscopy has revealed long-lasting oscillations
in the optical response of a variety of photosynthetic complexes. Different theoretical models which involve the coherent coupling of electronic (excitonic) or electronic-vibrational (vibronic) degrees of freedom have been put forward to explain these observations. The ensuing debate concerning the relevance of either one {\rm or} the other mechanism may have obscured their
potential synergy. To
illustrate this synergy, we quantify how the
excitonic delocalization in the LH2 unit of  {\it Rhodopseudomonas Acidophila} purple
bacterium, leads to correlations of excitonic energy fluctuations, relevant coherent vibronic coupling and, importantly, a decrease in the excitonic dephasing rates. Combining these effects, we identify a feasible origin for the long-lasting oscillations observed in fluorescent traces from time-delayed two-pulse single molecule experiments performed on this photosynthetic complex.

\end{abstract}

\maketitle

Long-lasting oscillations --ranging from  hundreds of femtoseconds at  room temperature to a few picoseconds at 77K--  in the non-linear optical response from the Fenna-Matthews-Olson (FMO) complex in green sulphur bacteria \cite{Engel_Nature2006,EngelNJP,Fleming2005},  purple bacteria  light harvesting complexes \cite{Engel_PNAS2012}, or of  reaction centers of bacteria and  higher plants\cite{Flemming_2007Science,Ogilvie_NChem2014,Romero_NPhys2014} have been reported. These observations are obtained by  excitation and read out with four calibrated ultra-short pulses providing rich but spectrally congested data of the  ensemble dynamics, with a complexity that has made difficult to reconcile some observations\cite{Engel_Nature2006,MIller_PNAS2017,Donatas_2017}.
The most promising models  to explain these experiments rely on the generation of protected excitonic coherence due to correlated fluctuations \cite{Flemming_2007Science,Ishizaki2010} or on the existence of long-lived  vibronic coherence induced and sustained by the interaction between excitons and underdamped vibrational modes \cite{PlenioNature2013,Christensson_JPCB2012,Chenu_2013SciRep,Plenio_2013JCP,Tiwari2013,Novelli_JPCLett2015,
Prior,Chin,HuelgaP13,ChinHuelgaPlenio2012,Womick2011}. 

Remarkably, these observations of long-lasting oscillations in photosynthetic structures have also found support from a different type of setup \cite{Hildner2013} which avoids ensemble averages and congested spectra, as it reports oscillating fluorescence traces from single light harvesting 2 (LH2) complexes from purple bacteria, as a function of the time delay between only two pulses, or as a function of the phase difference between pulses for a given time delay. Motivated by this experiment, in this article we provide a careful analysis of the subtleties of the excitonic delocalization in LH2 that leads to  a substantial and robust vibronic interaction,  which, in turn, identifies this vibronic-excitonic synergy as the possible origin of the observed oscillations.


\begin{figure}
\centering
\includegraphics[width=1.\columnwidth]{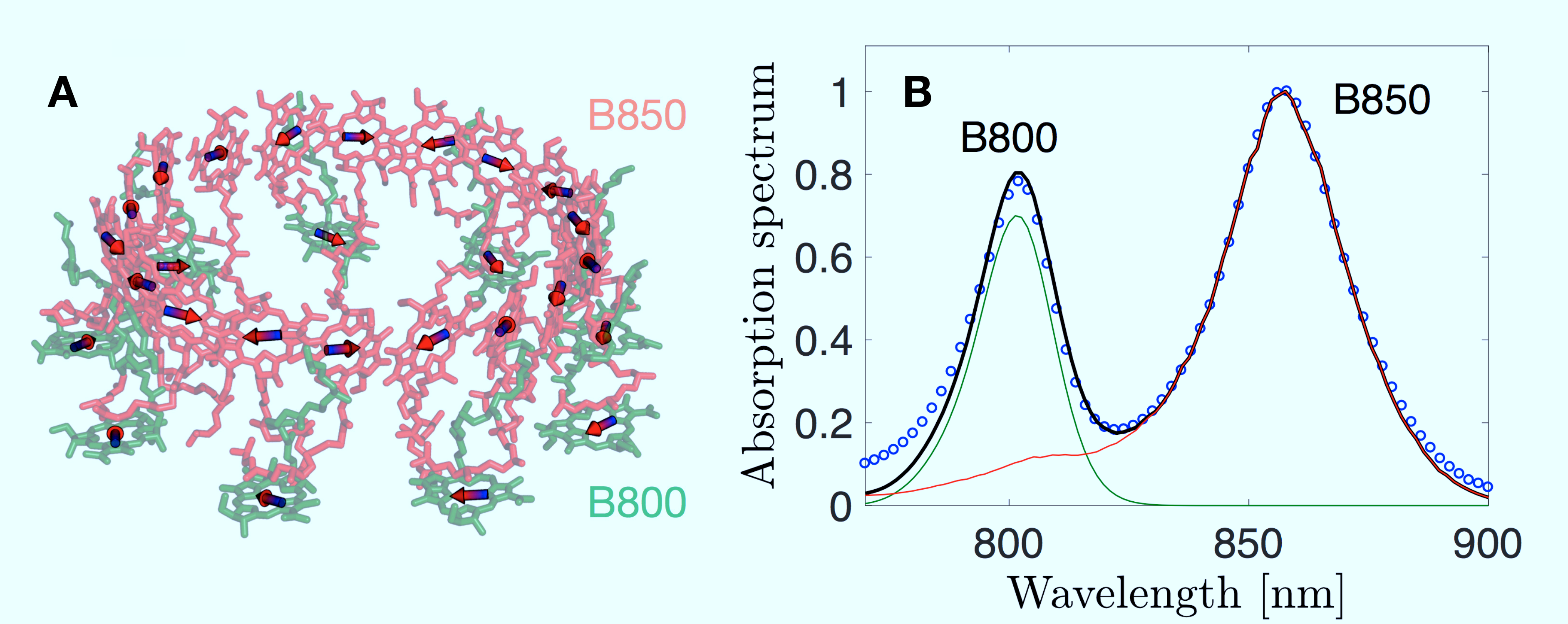}
\caption{{\bf A} The schematic representation  LH2 from {\it Rps. acidophila} based on its  X-ray structure \cite{PDBfile}. The inner and outer rings of pigments correspond to the B850 and B800 structures,  respectively, with $Q_y$ transition dipoles denoted by arrows.  {\bf B} Experimentally observed absorption of the LH2 complex \cite{Hildner2013} (blue circles) superimposed to our model calculation (thick black continuous). The spectra associated to  the excitons mainly delocalised over the B800 and B850 rings are shown in green and red lines, respectively.  Static disorder of B850 and B800 pigments is modelled by independent variations of pigment's energy according to Gaussian distributions with respective  standard deviations (SDs) of 350 cm$^{-1}$ and 150 cm$^{-1}$.
}
\label{fig1}
\end{figure}


High resolution X-ray crystallography of the LH2 complex of  {\it Rhodopseudomonas (Rps.) Acidophila} \cite{PDBfile} reveals a structure composed of  9-fold repeating apoproteins, monomeric in the B800 and dimeric in the B850 rings, with absorption maxima at about 802 and 858 nm respectively \cite{Codgell2006}, as shown in Fig.\ref{fig1}. Several theoretical and experimental
 studies  \cite{Jimenez_JPCB1997,vanGrondelle_CPL1997,Krueger1998,vanGrondelle_1998Biochem, Sundstrom,vanOIjeen_BioPhys2000,vanGrondelle_1997JPCB,Hu_2002,Trinkunas_JLum_2003,Timpmann2004, Urboniene_2005Pres,Silbey_PRL_2006,Engel_PNAS2012,vanGrondelle_PNAS2016}  
have reached a partial consensus on quantities relevant for excitonic dynamics, such as excitonic couplings, pigment's energies and their environment-induced fluctuations.  Hildner {\it et al.} \cite{Hildner2013} reported, unexpectedly, long-lived oscillations in the fluorescence intensity ($FI$) traces from two color pulsed illumination, in a scheme intended to first excite the B800 band and then read out the B850 pigments' population in single LH2 complexes with the second pulse. By varying the time-delay between pulses, oscillations in the $FI$ were observed up to 400\,{\rm fs} with a period of $\sim$\,200\,{\rm fs}~\cite{Hildner2013}. Oscillations with about the same  amplitudes were also observed  by varying the phase difference between pulses  at a fixed time delay.  
The oscillations were interpreted as signatures of coherent exciton exchange between B800 and B850 rings. Since the B800$\rightarrow $B850 process is regarded as an incoherent excitonic transfer \cite{Silbey_PRL_2006,Pullerits1997}, this result prompted a reevalution of the excitonic dephasing rates and/or an analysis of additional degrees of freedom that may be able to explain these oscillatory features. We will reexamine the coherent B800-B850 excitonic interaction in order to provide support for a different scenario  in which the excitons delocalised over  common pigments in the B850 ring participate in long-lasting coherent oscillatory population exchange mediated by underdamped vibrational modes. We will show that the delocalization of the bright and dark excitonic wavefunctions across B850 pigments fulfils the requirements of mode-mediated coherent population exchange, namely, strong vibronic coupling, slow excitonic dephasing, and mode-exciton resonance insensitive to static disorder. We will show that the analysis of the $FI$ traces as a function of the phase difference  between pulses further supports our model and, specifically, the significance of excitonic delocalization across the B850 structure.


The  $Q_y$ transition dipoles $\vec d_i$,  shown schematically in  Fig.\ref{fig1}{\bf A}, from the electronic ground state $\ket{g}$ to the  excited state $\ket{i}$ of the $i$th pigment, present  a mutual interaction, $J_{ij}$, described by the excitonic Hamiltonian
\bea
\mathcal{H}_e &=& \sum_{i}^{N_{800}+N_{850}} \Omega_{i}\ket{i}\bra{i} + \sum_{i\ne j}^{N_{800}+N_{850}}  J_{ij}\ket{i}\bra{j}\nonumber\\
&=&\sum_{\stackrel{\alpha_m\in \mbox{\tiny B800}}{ \beta_k\in \mbox{\tiny B850}}} \epsilon_{\alpha_m}\ket{\alpha_m}\bra{\alpha_m}+\epsilon_{\beta_k}\ket{\beta_k}\bra{\beta_k}\nonumber\\
&&+V_{\alpha_m,\beta_k}(\ket{\beta_k}\bra{\alpha_m}+\ket{\alpha_m}\bra{\beta_k})\nonumber\\
&=&\sum_k \epsilon_{\gamma_k}\ket{\gamma_k}\bra{\gamma_k}.\label{ElectronicHamiltonian}
\eea
The excitons $\ket{\alpha_k}=\sum_{i\in{\mbox \tiny B800}} c_i^{\alpha_k} \ket{i}$ and $\ket{\beta_k}=\sum_{i\in\mbox \tiny B850} c_i^{\beta_k} \ket{i}$ 
are the eigenstates within B800 and B850 rings, respectively, for the case when the interaction between B800 and B850 rings  is neglected ($V_{
\alpha,\beta}=0$). The excitons $\ket{\gamma_k}=\sum_{i\in {\mbox\tiny B800,  B850}}c_i^{\gamma_k} \ket{i}$ are the eigenstates of the full excitonic Hamiltonian. In agreement with previous studies \cite{Pullerits1997} we will obtain that due to the weak inter-ring B800-B850 coupling $|V_{\alpha,\beta}|\lesssim 30$ cm$^{-1}$ (see SI and \cite{Pullerits1997,Silbey_PRL_2006}), these eigenstates can be grouped into states mostly delocalised over the B850 or the B800 pigments, i.e., $\ket{\gamma_k}\approx \ket{\beta_k}$ or $\ket{\gamma_k}\approx \ket{\alpha_k}$, respectively, which will be called B850 and B800 states in this work.

Due to the presence of water molecules or protein residues, the pigment energies  $\Omega_i$ vary in a time-scale longer than the excitonic lifetime, in a process termed static disorder. In our description we assume   that the fluctuations of pigment energies are uncorrelated  $\langle\Omega_i\Omega_j\rangle= \langle \Omega_i\rangle\langle\Omega_j\rangle$, as expected from local  interactions affecting individual pigments \cite{Olbrich2011}.


The excitation by the pulses is described by the light-matter interaction Hamiltonian $H_{field}(t)=\vec E(t,\phi)\cdot(\sum_{k}\vec{D}_{\alpha_k}\ket{\alpha_k}\bra{g}+\sum_{\beta_k}\vec{D}_{\beta_k}\ket{\beta_k}\bra{g})+h.c.$ where $\vec E(t,\phi)=E(t,\phi)\hat E$ gathers the polarization $\hat E$, amplitude and phase $\phi$ of the  laser pulses. The transformed vectors of pigments' dipoles $\vec{d}_i$ to the exciton basis  $\vec{D}_{\alpha_k}$, $\vec{D}_{\beta_k}$  can lead to  dark excitonic states, which,   whenever their absorption cross section  is less than $1\%$ of the total  LH2 cross section (see SI) will be denoted by  $^*$.

 \begin{figure}
\centering
\includegraphics[width=1.\columnwidth]{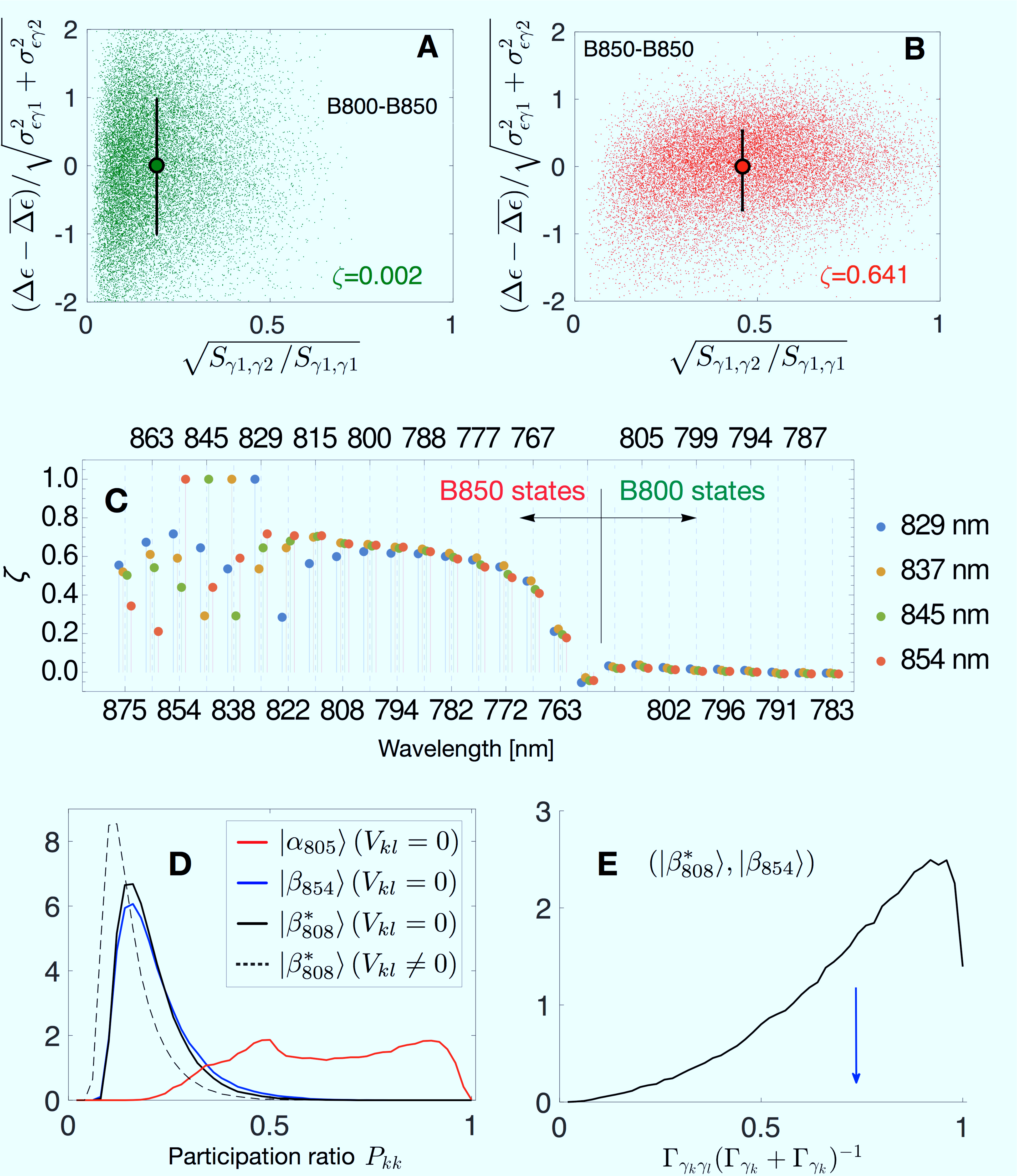}
\caption{ The effect of static disorder in vibronic coupling, excitonic splittings and excitonic dephasing in LH2. {\bf A} ({\bf B}) Distribution of excitonic splitting and the ratio of diagonal and off-diagonal vibronic couplings of $\ket{\gamma_{805}\approx\alpha_{805}}$ and $\ket{\gamma_{854}\approx\beta_{854}}$ states ($\ket{\gamma_{808}\approx\beta_{808}^{*}}$ and $\ket{\gamma_{854}\approx\beta_{854}}$ states). Note the larger amplitude of excitonic gap fluctuations for the B800-B850 pair in {\bf A} and the increased excitonic exchange coupling $S_{kl}$ for the B850 pair in {\bf B}.   The larger correlation coefficient $\zeta$ observed in {\bf B} 
is shown to persist between the B850 states $\ket{\gamma}\approx\ket{\beta_{829}}$,$\ket{\beta_{837}}$,$\ket{\beta_{845}}$, $\ket{\beta_{854}}$,  and all other B850 LH2 excitons, as is shown in {\bf C}. This supports the existence of correlated static disorder underpinned by excitonic delocalisation.  {\bf D} Distribution of the participation ratio $P_{kk}$ for the brightest states $\ket{\alpha_{805}}$, $\ket{\beta_{854}}$ of each ring, and the dark state $\ket{\beta_{808}^*}$ for coupled and uncoupled rings ($V_{\alpha,\beta}=0$ and $V_{\alpha,\beta}\ne 0$). {\bf E} Distribution of the ratio of excitonic dephasing rate and the sum of optical dephasing rates of $\ket{\beta_{808}^{*}}$ and $\ket{\beta_{854}}$ states. The arrow shows the average ratio $\simeq 0.7$, which due to a delocalisation over common pigments, as expressed in Eq.(3) and (4),  sets a reduced inter-excitonic dephasing as compared to the optical dephasing.  All the distributions were obtained with $10^5$ random realizations of pigment energies.}\label{fig2}
\end{figure}

The interaction between  the excited state of the pigment and a single vibration per pigment --with  energy  described by ${\mathcal H}_v= \sum_{i} \omega b_{i}^\dagger b_{i} $-- is modelled by a linear coupling term of the form
\begin{align}
\mathcal{H}_{e-v} &=  \sum_{i} \sqrt{s}\omega \left( b_{i}^\dagger +  b_{i} \right)\ket{i}\bra{i}\\
&=\sum_{k,l}  \sqrt{S_{kl}}\omega  \left(a_{kl} + a_{kl}^{\dagger}\right) \ket{\gamma_k}\bra{\gamma_l}\label{H2}
\end{align}
which quantifies  the vibration-pigment coupling strength through the  Huang-Rhys ($HR$) factors $s$. 
As explained in the SI,  we collect all modes of identical frequency $\omega$ coupled to different pigments  in eq.(\ref{H2}) , into  a generalised coordinate with creation (annihilation) operators $a_{kl}^\dagger$ ($a_{kl}$). This coordinate is a linear function of the operators $b_i^\dagger$ ($b_i$) and obeys the canonical commutation relations $[a_{kl},a_{kl}^\dagger]=1$  and the generalised $HR$ factor $S_{kl}=s\sum_{i}|\bracket{\gamma_k}{i}\bracket{i}{\gamma_l}|^{2}=s\, P_{kl}$.   The quantity $P_{kl}=\sum_i|\bracket{\gamma_k}{i}\bracket{i}{\gamma_l}|^2$ is the  spatial overlap of the excitonic wave-functions $\ket{\gamma_k}$ and $\ket{\gamma_l}$, while $P_{kk}^{-1}$ is the inverse participation ratio and quantifies the number of pigments that participate in exciton $k$. The diagonal $S_{kk}$ results in the  vibrational progression due to a redistribution of dipole strength on side-bands and overtones, whereas $S_{kl}$, with $k\ne l$, mediates the excitonic population exchange between $\ket{\gamma_k}$ and $\ket{\gamma_l}$. Notice that this coherent exchange occurs only if excitonic wave functions overlap. Moreover,  this population exchange is important if the resonance condition $\omega\approx \epsilon_{\gamma_k} -\epsilon_{\gamma_l}$ is met. 

Figure \ref{fig2}{\bf A} and {\bf B}  present scatter plots  of the variations  $\Delta E$ of the energy gap between specific pairs of excitons $\ket{\gamma_k}$ and $\ket{\gamma_l}$, with respect to their average energy gap $\overline{\Delta E}$. These figures consider  the brightest exciton of the B850 band, with either the brightest B800 exciton in  {\bf A}, or with a dark B850 state lying close to 800 nm in  {\bf B}. The variations presented are normalised with respect to the standard deviation (SD) of this gap if the static fluctuations of these excitons were uncorrelated 
 $\sigma_{ind}=\sqrt{\sigma_{\epsilon_k}^2+\sigma_{\epsilon_l}^2}$. Here, $\sigma_{\epsilon_k}$ represents the SD of the excitonic energy $\epsilon_k$. The SD  of excitonic energies of the B800 excitons are very similar to those of their pigments, while due to the important coupling between the B850 pigments, the  exchange narrowing \cite{VanGrondelleExcitonsBook} reduces the SD  $\sigma_{850}=350$ cm$^{-1}$ of pigment energies, to a SD of B850 excitons  $\sigma_{\beta_k}\approx 107$ cm$^{-1}$. The ratio $(\Delta E-\overline{\Delta E})/\sigma_{ind}$
will thereby have  a confidence interval of $\pm$1 SD which is equal to two for uncorrelated fluctuations.
Figure \ref{fig2}{\bf A} shows that the confidence interval for the B800-B850 excitonic energy gap is indeed equal to two. However,  the confidence interval of $\pm 1$SD for $(\Delta E-\overline{\Delta E})/\sigma_{ind}$  in  Fig.\ref{fig2}{\bf B} is smaller than two, implying that the energy gaps in the B850 excitons displayed here, present reduced variations than those expected from independent exchange narrowed transitions. The robustness of the excitonic gaps between B850 excitons can be associated to correlations of these fluctuations. The correlation of excitonic energies can be quantified by   the Pearson correlation coefficient $\zeta\equiv$$[\langle\epsilon_{k}\epsilon_{l}\rangle-\langle\epsilon_k\rangle\langle \epsilon_l\rangle]/(\sigma_{\epsilon_k}\sigma_{\epsilon_l})$. This coefficient has a value  $\zeta=0.002(8)$ for the energies of B800-B850 states in  Fig.\ref{fig2}{\bf A} contrasting with a large correlation of $\zeta=0.64(8)$ for the B850 states in {\bf B}. The origin of this correlation will be more transparent after determining the subtleties of the excitonic delocalisation across the LH2 pigments.

To that end, we also consider in Fig.\ref{fig2}{\bf A} and {\bf B}  the scatter plots of the excitonic overlap $P_{kl}/P_{kk}=\sqrt{S_{kl}/S_{kk}}$. Unsurprisingly, the overlap between B800 and B850 excitons in  {\bf A}  is two to three times smaller than the overlap  between B850 excitons in  {\bf B}. Since correlations of the excitonic gaps arise in  {\bf B} but not in  {\bf A}, these results  relate a larger overlap of the excitonic wavefunctions to a larger  correlation among their respective energy fluctuations.

Figure \ref{fig2}{\bf C} generalises the relation between excitonic overlap and correlations in the fluctuations. Here we can see that a high correlation of fluctuations between all B850 states, contrasts with a low correlation between B850-B800 exciton pairs.  We conclude that the correlations in excitonic energy variations are important for overlapping excitonic wavefunctions, implying that perturbations on individual pigment energies affect all excitonic energies that delocalise over that pigment, in a similar way. As a consequence of  the overlap of excitonic wavefunctions, the static disorder may compromise the resonance between a mode and an excitonic gap $\omega\simeq \epsilon_{\gamma_k}-\epsilon_{\gamma_l}$ in the B800-B850 exchange, while due to the correlations observed,  this resonance is more robust for the vibronic exchange between B850 excitons.

The explicit connection of this excitonic overlap with the ratio  between the non-diagonal and diagonal vibronic interaction, $\sqrt{S_{kl}/S_{kk}}=P_{kl}/P_{kk}$, described after  eq.(\ref{H2}), expresses the strong influence of the  excitonic overlap $P_{kl}$ on the coherent vibronic exchange of populations between these excitons. Then, the appreciable excitonic overlap in the B850 excitons provides robust inter-excitonic energy gaps, while at the same time provides an important vibronic coupling for excitonic population exchange.

Excitonic delocalization also affects the dephasing rates of the excitons, which determine the duration of the coherent population  exchange between excitons. The Redfield formalism \cite{Lim_2015NC,BreuerPetrucione}  specifies the inter-exciton  dephasing rate  in terms of the  pigment's  dephasing rate $\Gamma_i$  of their  ground-excited state coherence
\begin{equation}\label{eq:excitonic_dephasing}
\Gamma_{\gamma_k,\gamma_l}=\Gamma_{\gamma_k}+\Gamma_{\gamma_l}-2\sum_i|\bracket{\gamma_k}{i}\bracket{i}{\gamma_l}|^2\Gamma_i
\end{equation}
which depends on the optical coherence $\ket{g}\bra{\gamma_k}$ dephasing rate
\begin{equation}\label{eq:optical_dephasing}
 \Gamma_{\gamma_k}=\sum_{i}|\bracket{i}{\gamma_k}|^4 \Gamma_i+\frac{1}{2}\sum_{l\neq k}\Gamma_{k\rightarrow l},
\end{equation}
where $\Gamma_{k\rightarrow l}$ is the relaxation rate from $\ket{\gamma_k}$ to $\ket{\gamma_l}$ (see SI).   Two reductions of dephasing rates are discernible: in eq.(\ref{eq:excitonic_dephasing})  the inter-excitonic  dephasing rate is suppressed thanks to the overlap between excitonic wave-functions, while in eq.(\ref{eq:optical_dephasing}),  $\Gamma_{\gamma_k}\approx P_{kk}\Gamma_i$ is reduced due to the excitonic participation ratio.  Diagonalization of realizations of $\mathcal{H}_e$ under static disorder in Fig.\ref{fig2}{\bf D}, results in averages for B850 states of $P_{kk}^{-1}\approx 5$ nearly independent on whether  the B800-B850 coupling is  considered or not, further supporting the designation of B800 and B850 excitons. Previous results indicated a smaller delocalization,  however, associated to the much longer superradiance time-scale \cite{vanGrondelle_1997JPCB}.  The optical dephasing in the B850 excitons reduces {\it on average} to about one-fifth from that of individual pigments, while
the B800 excitons  limited to  $P_{kk}^{-1}\approx 1.5$ pigments, do not benefit from this reduction. Although this reduction is reminiscent of the exchange narrowing mechanism \cite{PhotoEx}, it is associated however, to excitonic decoherence and not to static disorder.

Figure \ref{fig2}{\bf E} displays the distribution of $\Gamma_{\gamma_k \gamma_l}(\Gamma_{\gamma_k}+\Gamma_{\gamma_l})^{-1}$ for the states $\ket{\beta_{808}^*}$ and $\ket{\beta_{854}}$  under static disorder. Even though this ratio has an average value of  $\approx 0.7$ (highlighted by the arrow) for this set of states, it must be underlined that there are many events with much lower dephasing, whose observation is accessible in single molecule experiments such as Ref. \cite{Hildner2013}.  After taking into consideration the relaxation within the B850 manifold and our estimate of $\Gamma_i=377$ cm$^{-1}$ (optical coherence decay in 30 fs) from experiments in B820 dimers \cite{vanGrondelle_CPL1997} (further details in the SI), we obtain  excitonic optical and  inter-excitonic  dephasing rates of $\Gamma_{\beta_k} \approx$74 cm$^{-1}$ and $\Gamma_{\beta_{854}, {\beta_{808}^*}}=106$ cm$^{-1}$, resulting in 140 fs  and 106 fs decay time constants, respectively. Inter-excitonic  B800-B850 coherence presents a decay rate of 1/60 fs which is almost twice as fast as the decay of inter-excitonic  B850 coherence. Notice that beyond average values, the distribution in Fig.\ref{fig2}{\bf E} indicates that a dephasing slower than, e.g., 200 fs, is expected for  14\% of the traces. Observe that this is already about 7 times the lifetime of optical coherences from individual pigments.

 \begin{figure}
\centering
\includegraphics[width=1.01\columnwidth]{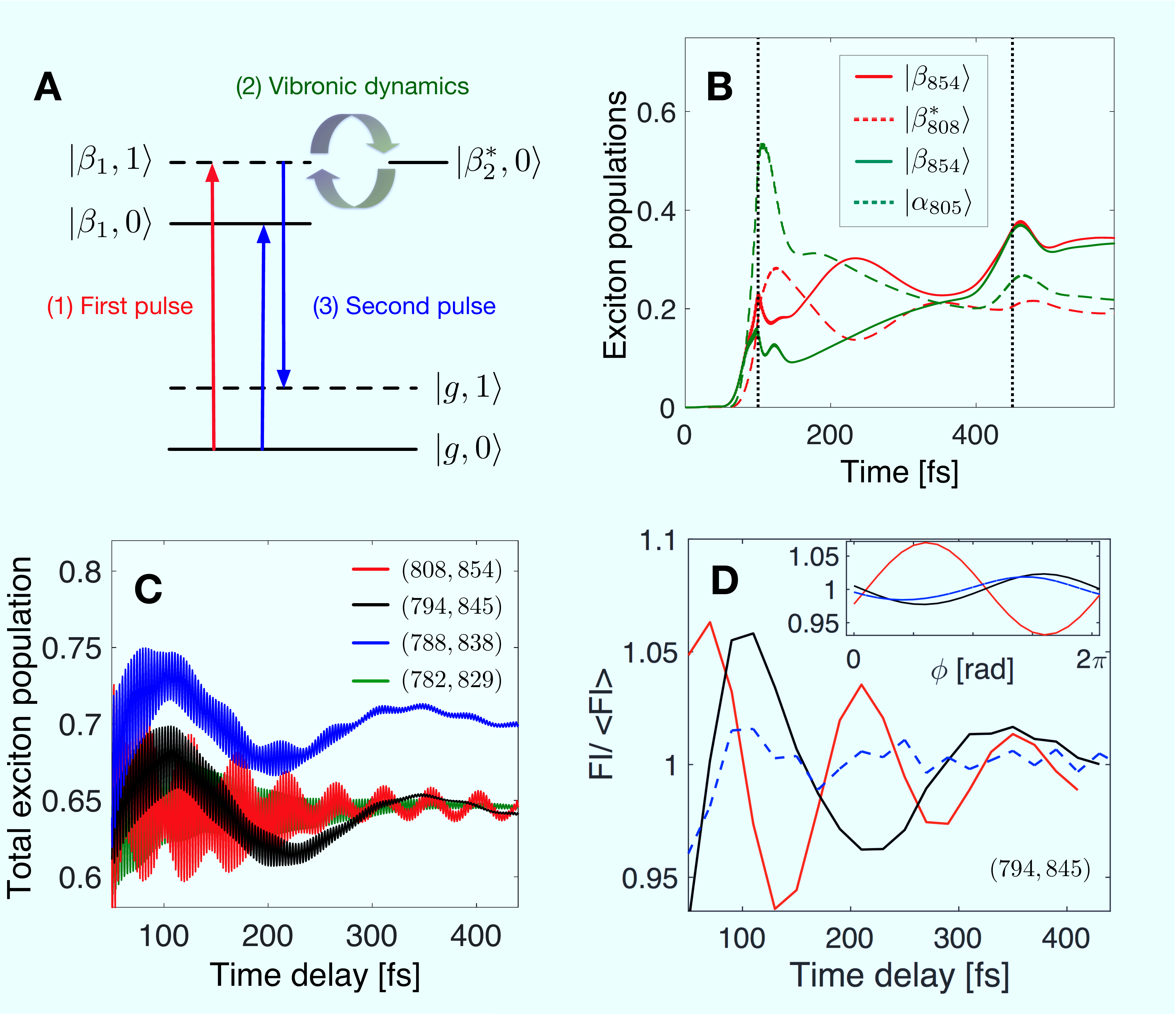}
\caption{{\bf A} Schematics of a vibronic model: The first pulse (red) centered at around 800 nm excites the phonon sideband of a bright B850 exciton $\ket{\beta_1,1}$, and the second pulse (blue) centered at around 850 nm de-excites the phonon sideband $\ket{\beta_1,1}\rightarrow \ket{g,1}$. The probability of the second transition is modulated by the population exchange between the phonon sideband and a dark B850 state$\ket{\beta_2,n=0}$, mediated by the vibronic coupling. {\bf B} Result of the populations along the pulse sequence (dotted lines represent the times at which pulses present maximum amplitude) for the B800-B850 vibronic model.  {\bf C} Populations from   pairs of excitons  which couple via a single vibrational coordinate and present an important optical response. Notice that the two pairs (794,845) and (788, 838) present traces with the observed $\simeq$200 fs period \cite{Hildner2013}. {\bf D} shows the total excited state population for the pair 845-794 nm normalised by its average $FI(T)/\langle FI\rangle$ for three cases; case 1: with vibronic interaction (black), case 2: without vibronic interaction (dotted, blue).  We present in red the expected signal for a different realisation of case 1, showing a different phase for the $FI$ trace. This different phase is consistent with the observations in \cite{Hildner2013}, that result in a non-oscillatory trace for the ensemble average.  Here the resolution of  time delays is $\Delta T= 20$ fs. The inset presents the total excited population $FI(\phi)$ as a function of the relative pulses phase $\phi$ for a delay time $T=100$ fs. A single mode with frequency 727 cm$^{-1}$ has been used in all cases.}
\label{fig3}
\end{figure}

We have discussed how the excitonic delocalization affects  i) the robustness of the vibronic resonance, ii) the magnitude of the vibronic coupling, and iii) the inter-excitonic dephasing rates.  
We have chosen the excitons in Fig.\ref{fig2}{\bf A} and {\bf B} such that their energy difference is on average similar to the central frequency difference between the laser pulses used in Ref. \cite{Hildner2013} . This condition ensures that similar pairs as those in Fig.\ref{fig2}{\bf B} present a side band of the B850 bright state which becomes excited by the first pulse.   Due to the robustness of the energy gap and the low inter-excitonic dephasing, a resonant long-lasting  population  exchange could proceed with the zero-phonon line of the dark B850 exciton, lying at about 800 nm, via a vibronic mode. 
Hence, our proposal for explaining the observed oscillations assumes that the first pulse with a carrier wavelength of $\approx  800$ nm,  excites the phonon sideband of the bright B850 excitons, which in Fig.\ref{fig3}{\bf A} is labelled $\ket{\beta_1,1}$. The optical transition from the ground state $\ket{g,0}$ to this side-band is resonant with the side-band associated to modes with frequencies in the range of 700-900 cm$^{-1}$. Modes within this range were observed in LH2 \cite{Reddy1991} and in isolated BChl pigments \cite{Renge_1987JLum}, with frequencies 727 and 773 cm$^{-1}$
  and $S_{kk}=0.05$ \cite{Reddy1991}. These modes are relevant for the theoretical estimation of the B800$\rightarrow$ B850  transfer rate \cite{Pullerits1997}. Based on the averages from Fig.\ref{fig2}, if $S_{kk}=0.05$ as for these modes \cite{Reddy1991}, then $S_{kl}=0.002$ for the B800-B850 pairs of states, while a six times larger  $S_{kl}=0.012$ results for the pairs of B850 states. This results in a non-diagonal vibronic coupling strength  $\omega\sqrt{S_{kl}}$ of approximately $\approx 80$ cm$^{-1}$ for the B850 excitons, which would account for oscillations of populations between B850 excitons having a period of about  200 fs. The observation of several $FI$ traces with periods commensurate to this latter value \cite{Hildner2013}, suggests that the vibronic couplings between B850 excitons (and not between B800 and B850 excitons) is a good candidate for the origin of these oscillating traces. As presented schematically in Fig.\ref{fig3}{\bf A}, according to the Hamiltonian eq.(\ref{H2})  this non-diagonal vibronic interaction between B850 states induces coherent energy exchange linking the bright state side-band $\ket{\beta_1,1}$ and a dark state  $\ket{\beta_{2}^{*},0}$ without vibrational excitations, {\it e.g.}~$\ket{\beta_1}=\ket{\beta_{854}}$ and $\ket{\beta_{2}^{*}}=\ket{\beta_{808}^{*}}$. The second pulse mainly induces a population inversion of the zero-phonon transition between  $\ket{\beta_1,1}$ and $\ket{g,1}$ concomitant with the excitation from $\ket{g,0}$ to $\ket{\beta_1,0}$. The fluorescence intensity is proportional to the total excited population after the interaction with the second pulse.

In the following, we  demonstrate  numerically that vibronic dynamics within the B850 excitons can induce long-lasting oscillations in the $FI$ traces with $\sim 200\,{\rm fs}$ period. To take into account the influence of laser pulses on the $FI$ dynamics  accurately, we consider the experimental laser spectrum and apply a linear phase below 820\,nm, as performed in Ref.\cite{Hildner2013},  in order to generate two pulses (the first with an almost flat spectrum between 780 and 820\,nm; the second peaking at 828 nm) with a controllable time-delay $T$, which is determined by the slope of the linear phase in frequency domain (see the SI).

Accordingly, the models we present in Fig.\ref{fig3} consist of a pair of excitons with the wavelengths and transition dipoles obtained from averages over static disorder (see Fig 4-{\bf C} in the SI), which couple via a single mode, that represents a generalised coordinate involving the vibrations at each pigment, having a frequency of 727 cm$^{-1}$. Notice however, that these pigment modes resonantly  couple several bright states whose side-band can be excited by the first pulse (four excitons with zero-phonon lines lying between 854 and 829 nm), with several  complementary  dark excitons (between 808 and 782 nm). Hence, the rather stringent vibronic resonance condition is, on average, fulfilled by multiple excitons, and, as here underlined, robust to static disorder due to overlapping excitonic wavefunctions. The evalution of the dynamics of these excitons  coupled by multiple vibrational modes, plus the exciting fields probing multiple delay times, is not amenable for computational purposes. Hence, we study the qualitative features of signals arising from selected pairs of B850 excitons and a single mode of frequency 727 cm$^{-1}$ that couples them.
In Fig.\ref{fig3}{\bf B}, the population dynamics of two cases ($\ket{\gamma_k\approx\beta_{808}^{*}}$, $\ket{\gamma_l\approx\beta_{854}}$) and ($\ket{\gamma_{k}\approx\alpha_{805}}$, $\ket{\gamma_l\approx\beta_{854}}$), are shown, where the B850 states in the former case exhibit oscillatory population dynamics during $T$ with a period of $\sim 200\,$fs (red) while the pair of B800 and B850 states does not (green). It is important to highlight that the  pulse intensity used in Ref.\cite{Hildner2013} is high enough to enable an appreciable excitation and inversion of the side-band $\ket{\beta_1,1}$ ensuing  a noticeable amplitude for the oscillations in the $FI$ traces.

 We note that all the $FI$ transients show ultrafast oscillations on top of slow dynamics on 100-200\,fs timescales. An additional component with period $\simeq$ 45 fs  is observed on the $\{\ket{\beta_{808}^{*}}$, $\ket{\beta_{854}}\}$ pair due to the detuning with the second pulse carrier wavelength of $\simeq 830$ nm. These ultrafast oscillations are a result of  the optical coherences created by the first pulse (see the SI) and were not resolved in the experiment \cite{Hildner2013}, since traces were recorded with resolution of time delays   $\Delta T\approx 20\,{\rm fs}$. In Fig.\ref{fig3}{\bf D}  we present the total excited state population dynamics regarding the vibronic $\{\ket{\beta_{794}^{*}}$, $\ket{\beta_{845}}\}$ pair,
 with resolution   $\Delta T=20$ fs, in order to highlight that the experiment was sensitive to the slower  component with $\sim 200\,{\rm fs}$ period. This result, namely case 1 (black), is compared with case 2 (dashed): the same excitons but no vibronic interaction, and case 3  (red):  the same excitons and mode as in case 1, but the exciton $\ket{\beta_{845}}$ is shifted to a higher energy by a single SD of $\Delta\epsilon_{\beta_k,\beta_l}$ (cf. Fig.~\ref{fig2}{\bf B}), and the dynamics proceed with a slower optical dephasing decay $\Gamma_{\beta}=37$cm $^{-1}$. This optical dephasing is half of that from case 1, hence with a decay time of 280 fs. 

Notice that case 2 does not show oscillations because in our model, the vibronic interaction is required to coherently cycle excitonic populations and generate oscillations in  $FI(T)$. A comparison between case 1 and  3 illustrates how $FI$ changes  among traces taken from a single complex subject to static disorder. 
The higher frequency of oscillations in case 3 is induced by the increased detuning between mode frequency and exciton splitting. In case 3, the interplay between lower dephasing and larger detuning results in oscillation amplitudes comparable to the case 1 with higher dephasing and smaller detuning. 
As mentioned before, the fast  oscillating component in $FI(T)$ at optical frequencies  depends on the optical coherences produced by the first pulse. The exciting fields can imprint their relative phase $\phi$  in these optical coherences, and as a result, a modulation of $FI(\phi)$  given a fixed time delay $T$. This modulation is shown  in the inset of Fig.\ref{fig3}{\bf D} for the three cases of the main figure. As in the experiment \cite{Hildner2013}, $FI(\phi)$ presents a full $2\pi$ cycle. However, for cases 1 and 2 these oscillations result in  a smaller amplitude than $FI(T)$, while in case 3, the oscillations of $FI(T)$ and $FI(\phi)$ have  comparable amplitudes. Since the experiment \cite{Hildner2013} presented similar amplitudes for delay-time and phase dependent traces, just as case 3,  it is very likely that the experiment \cite{Hildner2013} accessed realizations with a favourable dephasing.  Notice that the different phases of the oscillations of $FI(T)$ and $FI(\phi)$, support the observation \cite{Hildner2013} that on average,  the oscillatory component of these traces  vanish.

Our vibronic model also results in the absence of oscillating $FI(T)$ traces from control experiments  \cite{Hildner2013} devised to only  excite  the 800 nm or the 830 nm spectral window (see SI for detailed discussion).

Based upon the qualitative agreement of our model with the experimental observations, we conclude that the coherent electronic dynamics of the LH2 complex observed in experiments may originate from the synergy between coherent vibronic dynamics and excitonic delocalization, this latter supporting an appreciable vibronic coupling, fixing a robust resonant interaction while suppressing excitonic decoherence rates. This cooperativity serves to illustrate that the prospect of  generalizing  vibronics as the cause of oscillating features in optical traces, does not only rely on modes which resonantly  couple to specific excitonic transitions, but on the subtleties  of the excitonic wavefunctions involved. In particular the observations of Hildner, {\it et al.} \cite{Hildner2013}, are compatible with a reduced excitonic dephasing that results in the modulation of $FI(\phi)$, and a robust  vibronic interaction that results in the oscillatory  traces $FI(T)$. Single molecule experiments are therefore a promising tool to unravel the interplay between excitonic delocalisation, static disorder and dephasing in the optical response of light-harvesting systems.

This work was supported by an Alexander von Humboldt Professorship, the ERC Synergy grant BioQ, and the EU STREP  QUCHIP. This publication
was made possible through the support of a grant from the
John Templeton Foundation. Santiago Oviedo aknoweledges MINECO FEDER funds FIS2015-69512-R and Fundaci\'on S\'eneca (Murcia, Spain) Project No. ENE2016-79282-C5-5-R. Niek F. van Hulst acknowledges funding by the ERC Advanced Grant 670949-LightNet, MINECO grants SEV2015-0522 and FIS2015-69258-P, Centres de Recerca de Catalunya (CERCA) and Fundaci\'o CELLEX (Barcelona). We acknowledge Richard Hildner for facilitating the experimental traces and laser spectra.

\clearpage

\begin{center}\Large{Supporting Information\\ On the theory of excitonic delocalization for robust vibronic dynamics in LH2}\end{center}

\appendix

\section{Parameters for the calculations and numerical simulations}

For reference and reproducibility we include the parameters employed throughout the main text and the supplementary information. The electronic couplings among the pigments were estimated based on a 
two-ring arrangement with dipole position and orientations taken from the crystallographic structure of the LH2, according to the dipole-dipole interaction.

For B800 chromophores: 

\begin{center}
\begin{tabular}{|c|c|c|}
 Parameter & Value & Description \\
 \hline
 $d_{800}$ & 6.5 Debye & Pigments transition dipole \\
 $\Omega_{800}$ & 12460 cm$^-1$ & $Q_y$ transition energy \\
 $\sigma_{800}$ & 150 cm$^-1$ & Site energy disorder \\
 $\Gamma_{i\in B800}$ & 140 cm$^-1$ & Monomer dephasing rate \\
 $\Gamma_{800}$ & 96 cm$^-1$ & Exciton dephasing rate \\
\end{tabular}\\
\end{center}

For B850 chromophores:

\begin{center}
\begin{tabular}{|c|c|c|}
 Parameter & Value & Description \\
 \hline
 $d_{850}$ & 6.4 Debye & Pigments transition dipole \\
 $\Omega_{850}$ & 12530 cm$^-1$ & $Q_y$ transition energy \\
 $\sigma_{850}$ & 350 cm$^-1$ & Site energy disorder \\
 $\Gamma_{i\in B850}$ & 373 cm$^-1$ & Monomer dephasing rate \\
 $\Gamma_{850}$ & 74 cm$^-1$ & Exciton dephasing rate \\
\end{tabular}
 \end{center}

\section{Exciton dynamics in LH2 complex of {\it Rps. Acidophila}}

\subsection{Characterization of excitonic degrees of freedom.}

\begin{figure}
\centering
\begin{minipage}{1\columnwidth}
\includegraphics[width=1\columnwidth]{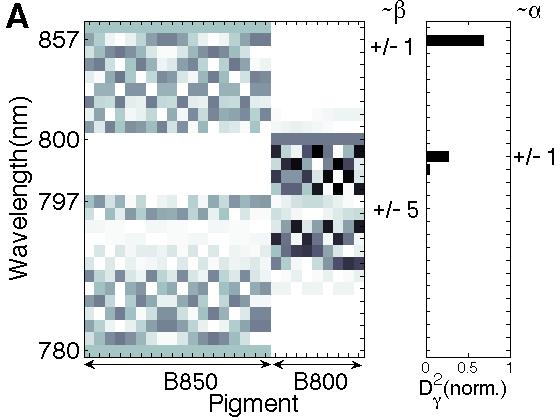}
\end{minipage}
\begin{minipage}{1\columnwidth}
\includegraphics[width=1\columnwidth]{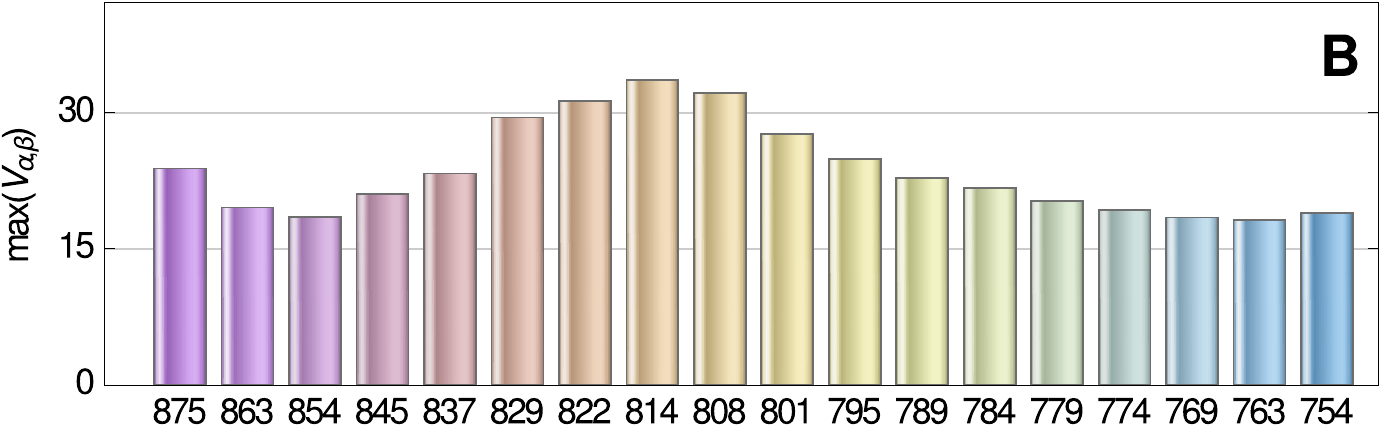}\\
\includegraphics[width=1\columnwidth]{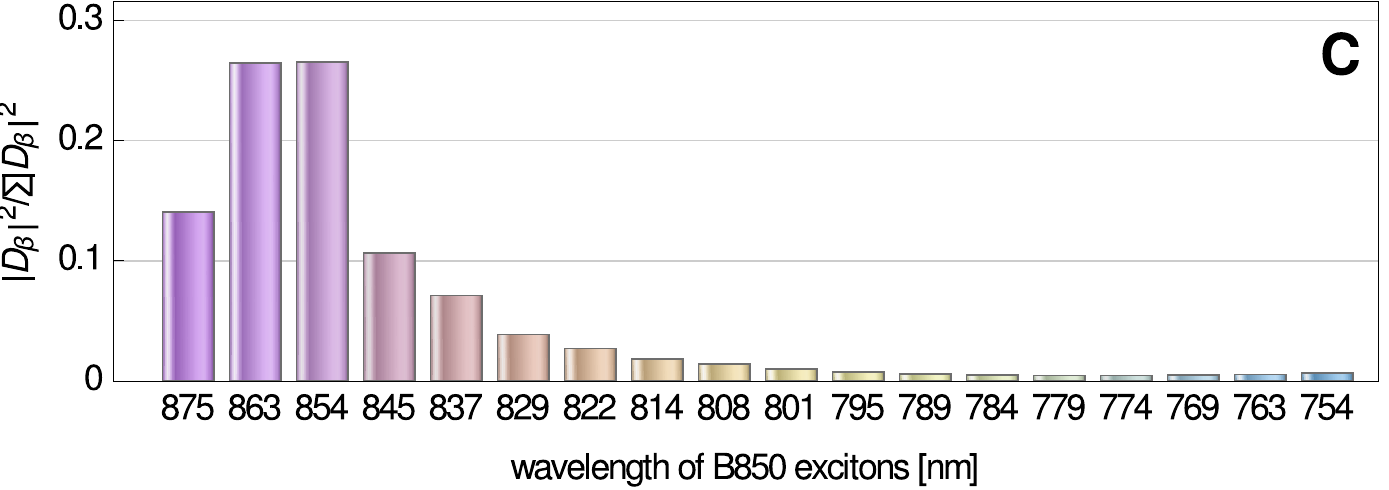}
\end{minipage}
\caption{ Characterization of excitons, B800-B850 coupling strengths and dipole strengths. In {\bf A} color map of pigment's populations of the full excitonic manifold of the LH2 in absence of 
static noise.  {\bf B}  Average of the maximum coupling between all the B800 states max$_{\alpha\epsilon B800} V_{\alpha,\beta}$ and each B850 state. {\bf C}  Average dipole strength of the 
B850 excitons $|D_\beta|^2$ as a percent of the B850 ring cross section $\sum_\beta |D_\beta| ^2$.  Averages from $10^5$ stochastic variations of B850 and B800 pigments' energies with respective 
variances  $\sigma_\omega=$350 and 140 cm$^{-1}$.}\label{fig1A}
\end{figure}

Diagonalization of the full electronic Hamiltonian in eq. (2) on the main text leads to the formation of 27 exciton states $\ket{\gamma}$, whose delocalisation domain is depicted in the colour map on 
Fig.\ref{fig1A}{\bf A}.  Of these states, 18 states $\ket{\beta}$ composing the B850 manifold and the remaining 9 states $\ket{\alpha}$ corresponding mostly to the B800. We observe that, in the case of a fully symmetric structure, where possible pigment's energies variations are neglected, exciton states delocalize mostly over just one 
ring.  The side panel of  Fig.\ref{fig1A}{\bf A} shows the induced 
transition dipole moment $\vec D_{\gamma}=\sum_i \bracket{\gamma}{i} \vec{d}_i$ for excitons $\ket{\gamma}$, These states ($\ket{\alpha = \pm 1}$ and $\ket{\beta = \pm 1}$ coincide with those obtained when  the 
inter-ring coupling $V$ is typically neglected\cite{Silbey_PRL_2006}. On the other hand, the maximum B800-B850 coupling (max($V_{\alpha,\beta}=\bra{\alpha} V\ket{\beta}$)) 
occurs between higher energy states located close to the B800 bright band, as  is shown in Fig.\ref{fig1A}{\bf B}. Besides a minor delocalisation between  these higher lying states of the B850 band and the B800 states, all other states can be safely labeled as either B800 or B850 excitons.

The optical response of the system is characterized by the
induced transition dipole moments associated to each exciton
state. In absence of any other coupling or perturbation, and for transition dipoles that present the tangential geometry along a circumference as shown in Fig.1{\bf A} in the main text,  the  pair $\ket{\beta=\pm 1}$ concentrate almost all the transition dipole of the full ring. Hence, a single band of the full ring arises peaking  at 850 nm, as shown in Fig.1 {\bf B} in the main text. In the prescence of static fluctuations, this dipole strength smears out across the B850 manifold, and results in the dipole strengths shown in Fig.\ref{fig1A}{\bf C}.

\subsection{Interaction with the environment}\label{Int_env}

Here we will characterize the influence of the environment on exciton dynamics, including decoherence, disorder and vibronic couplings.


\subsubsection{Exciton delocalization and dynamical properties}\label{char_excs}
The influence of an environment on the system dynamics is determined
by the coupling strength of each pigment to vibrations, characterized by phonon spectral density, and
electronic couplings between pigments, which lead to exciton
delocalization. A combination of  these elements with
experimental data is necessary to gain some insight on the appropriate
rates to model the complete dynamics of the LH2
system

The influence of environments on system dynamics is determined by the coupling strength of each pigment to vibrational environments, characterized by phonon spectral density, and electronic couplings 
between pigments, which lead to exciton delocalization. A combination of all this together with experimental data is necessary to gain some insight on the appropriate rates to model the complete 
dynamics of the LH2 system.

For the B850 band, slow dynamics of the nuclear relaxation rate \cite{Urboniene_2005Pres} underlies our use of Gaussian homogeneous line-shape functions 
(justified in the case of strong e-ph coupling \cite{Mukamel}) 
with a full-width at half maximum (FWHM) of  84 cm$^{-1}$. This value arises, as commented in the main text, from a 74 $cm^{-1}$ dephasing  component and a 20 cm$^{-1}$ intra-ring relaxation. The relaxation rate was estimated by a model of the full LH1 core complex\cite{Caycedo_JPC2017} and compared to experimental data of population depletion of the bright band in LH1 and LH2 complexes with a pump-dump-probe scheme \cite{Cohen_2011BioJ}. The inverse rate of 250 fs for the equilibration observed, results in  a FWHM of the Gaussian with a rate of 20 cm$^{-1}$. In order to 
get the 74 $cm^{-1}$ dephasing, an 
estimate of the pigments' dephasing rate $\Gamma_i$ for the B850 ring must be made. 
The relation between excitonic dephasing and pigment's dephasing derived in eqs.(4) and (5) in the main text requires a reliable estimation of $\Gamma_i$ for these pigments, which is --to our knowledge-- not 
available. The most simple system from which this dephasing rate rate can be estimated is the B820 dimer, believed to be composed of a dimeric unit of the B850 ring. In this subunit three-pulse 
photon echo peak shift has been performed, resulting in an amplitude of 100 cm$^{-1}$ and a time constant of 60 fs for the decay   of the excitonic  transition frequency correlation function 
\cite{vanGrondelle_CPL1997}. We use the cumulant expansion technique \cite{Mukamel,Cumulant}  to 
estimate a dephasing component of FWHM $\Gamma_{dimer}=$235 cm$^{-1}$, that is compatible with the calculation performed in LH1 complexes for the same decay time constant and amplitude using the same 
technique \cite{Jimenez_JPCB1997}.    Next, we assume that the pigments in B820 dimer are identical, and based in eqs.~(4) and (5) in the main text, we calculate $\overline{PR}_{\beta\, dimer}=\overline{\sum_i 
|\bracket{\beta}{i}^4}=0.61$ from simulation of the dimer energies including static disorder with an initial standard deviation $\sigma_\omega= 300$ cm$^{-1}$, that results in $\Gamma_i=385$ 
cm$^{-1}$. Using the 
same eq.~(5) in the main text, now applied to the full B850 ring, we estimate $PR_\beta=0.16$, resulting in an excitonic dephasing rate of $\Gamma_\beta=63$ cm$^{-1}$. This homogeneous broadening (plus 
the 20 cm$^{-1}$ associated to relaxation) results in a too narrow absorption spectra.
Greater static noise cause greater participation ratios and therefore, greater homogeneous line-widths. For $\sigma_\omega=350$ cm$^{-1}$, we obtain $PR_{dimer}=0.63$, 
$\Gamma_i=373$ cm$^{-1}$, $PR_\beta=0.2$,  $\Gamma_\beta=74$ cm$^{-1}$ and quantitative agreement with the experimental  absorption spectrum. 
The B800 pigments  are weakly coupled to phonon modes \cite{vanOIjeen_BioPhys2000,Katelaars_Biophys2001}, which explains  the narrower B800 (in comparison to the B850) absorption peak in Fig. 1{\bf 
B} 
in the main text. This fact was captured in our simulations by  inhomogeneous $\sigma_\omega$=150 cm$^{-1}$ and homogeneous $\Gamma_i= 140$ cm$^{-1}$ broadenings of the B800 pigments, which are 40\%  
from $\sigma_\omega$  and $\Gamma_i$ describing  the B850 pigments. For the B800 ring we obtained $\overline{PR_\alpha}=0.68$, thereby $\Gamma_{\alpha}=96$ cm$^{-1}$, which are also in good agreement 
with the absorption spectrum experiments.

We test the hypothesis of B800-B850 population exchange via excitonic degrees of freedom studying the coupling $V_{\alpha,\beta}$ under the influence of these inhomogeneities. As it is shown 
in Fig.\ref{fig1A}{\bf B},  the maximum excitonic coupling to the B800 excitons max($V_{\alpha,\beta}=\bra{\alpha} V\ket{\beta}$)  is not that from the brightest B850 states located at 862 and 854 
nm, but rather from  B850 states of higher energy at $\approx 805-820$ nm. Since these states lie close to the B800 excitons, the average coupling strengths of $\cong$ 40 cm$^{-1}$, can result in specific 
realizations in which excitons extend  over both B800 and B850 structures, analogous to the states $ \ket{\gamma\approx\beta=\pm5}$ highlighted in Fig.\ref{fig1A}{\bf A}. The average B850-B800 
couplings follow the same trend as the noise-less (no inhomogeneities) case, but smeared out and thereby with less variance:  the coupling decreases from 67 for $\ket{\beta=\pm 4}$ in the static case to 36 
cm$^{-1}$ in average, for the maximum coupling between B800  and the 813 nm state, but increases from 16 cm$^{-1}$ for  $\ket{\beta=0}$ to 24 cm$^{-1}$ for the lowest energy 874 nm state. The 
comparison of noise-less and average states is made based upon matching the ordering of excitonic energies in either case.

The coupling of the pigments with an environment  leads to a modification of the optical properties of the system, as the dipole strength is now redistributed among several excitons.
For the case of the LH2 complex, the average dipole strength of the B850 excitons in Fig.\ref{fig1A} {\bf C} shows that, the pair of states 
$\ket{\beta}$ at 863 and 854 nm  decrease each their dipole strength   to 26\%, measured as a percentage  $|\vec{D}_{\beta}|^2/\sum_\beta |\vec{D}_{\beta}|^2$ of the total B850 dipole strength 
$\sum_\beta 
|\vec{D}_{\beta}|^2$. The lowest energy state at 875 nm gains up to 14\%, while the states at 845, 837 and 829 nm, follow with a participation of 10\%, 7\% and 4\% of the total B850 absorption cross 
section. An analogous calculation  for the B800 ring results in a similar spread of dipole strength, however more even around the vicinity of the 
${\ket\alpha=\pm1}$, summarized by a participation in the total B800 ring cross section of 15, 17, 16, 13 and 10 percent for states with average wavelength of 812, 806, 804, 799 and 796 nm, 
respectively.

\subsubsection{Dephasing noise model}

In addition to static disorder, the optical response of molecules is affected by dephasing noise induced by the interaction between electronic and vibrational degrees of freedom. In this work, we consider local phonon environments where each pigment is coupled to an independent phonon environment modelled by a harmonic oscillator bath that is initially in a thermal state. The interaction Hamiltonian is described by
\begin{align}
	H_{e-ph}=\sum_{i}\ket{i}\bra{i}\otimes\sum_{\xi}g_{\xi}(B_{i,\xi}+B_{i,\xi}^{\dagger}),
\end{align}
where $B_{i,\xi}$ and $B_{i,\xi}^{\dagger}$ denote annihilation and creation operators, respectively, of a phonon mode with frequency $\omega_{\xi}$, determining the phonon spectral density ${\cal J}(\omega)=\sum_{\xi}g_{\xi}^{2}\delta(\omega-\omega_\xi)$. We assume that exciton dynamics is governed by master equation of Lindblad form where decoherence rates are characterized by the phonon spectral density and excitonic delocalization over sites \cite{BreuerPetrucione,Lim_2015NC}. The exact form of the factors that determine decoherence rates stems from representing the interaction Hamiltonian in the exciton basis $\ket{\gamma_k}=\sum_{i}\ket{i}\langle i|\gamma_k\rangle$, which will be summarized below.

Within the Lindblad description, the pure dephasing noise, which destroys coherences without inducing population relaxation in the exciton basis, is described by
\begin{align}
	L_{pd}=\sum_{i}\Gamma_{i}(0)\left(A_{i}(0)\rho(t)A_{i}(0)^{\dagger}-\frac{1}{2}\{A_{i}(0)^{\dagger}A_{i}(0),\rho(t)\}\right),
\end{align}
where $\Gamma_{i}(0)=\lim_{\omega\rightarrow 0}2\pi{\cal J}(\omega)(n(\omega)+1)$ is characterized by the phonon spectral density ${\cal J}(\omega)$ of site $i$ and mean phonon number $n(\omega)=1/(\exp(\hbar\omega/k_B T)-1)$ at temperature $T$ \cite{RivasHuelga_2010}. The operators $A_{i}(0)=\sum_{k}|\langle i|\gamma_k\rangle|^{2}\ket{\gamma_k}\bra{\gamma_k}$, originating from the interaction between sites and their local phonon environments, lead to modified dephasing rates for the excitonic states, depending on how exciton states $\ket{\gamma_{k}}$ are delocalized in the site basis $\ket{i}$. In the case of a two-level monomer, the dephasing rate of optical coherence, determining homogeneous broadening, is fully characterized by the pure dephasing rate $\Gamma_i\equiv \Gamma_i(0)/2$, as there is one excited state and therefore one can ignore the relaxation process amongst excited states, which is relevant only in multi-pigment networks: within the formalism, exciton relaxation is described by
\begin{align}
	L_{rxn}=\sum_{\omega\neq 0}\sum_{i}\Gamma_{i}(\omega)\left(A_{i}(\omega)\rho(t)A_{i}(\omega)^{\dagger}-\frac{1}{2}\{A_{i}(\omega)^{\dagger}A_{i}(\omega),\rho(t)\}\right),
\end{align}
where $\omega=\Delta\epsilon_{kl}=\epsilon_{\gamma_k}-\epsilon_{\gamma_l}\neq 0$ denotes the energy difference between exciton states $\ket{\gamma_k}$ and $\ket{\gamma_l}$, and $A_{i}(\omega)=\sum_{k,l}\langle \gamma_l | i\rangle\langle i|\gamma_k\rangle\ket{\gamma_l}\bra{\gamma_k}$.

For the Lindblad model, one can show that the dephasing rates $\Gamma_{\gamma_k}$ of optical coherences between ground and exciton states $\ket{g}\bra{\gamma_k}$ are given by \cite{Lim_2015NC,1367-2630-9-3-079}
\begin{align}
	\Gamma_{\gamma_k}=&\frac{1}{2}\sum_{i}|\langle i| \gamma_k\rangle |^{4}\Gamma_{i}(0)+\frac{1}{2}\sum_{l\neq k}\Gamma_{k\rightarrow l} \nonumber\\
\equiv& \sum_{i}|\langle i| \gamma_k\rangle |^{4}\Gamma_{i}+\frac{1}{2}\sum_{l\neq k}\Gamma_{k\rightarrow l},
\end{align}
where the contribution of the pure dephasing noise, characterized by $\Gamma_{i}(0)$, is represented as a function of the dephasing rate of a two-level monomer, $\Gamma_i =\Gamma_{i}(0)/2$, as shown in the main text, and the relaxation rate $\Gamma_{k\rightarrow l}$ from exciton $\ket{\gamma_k}$ to the other exciton state $\ket{\gamma_l}$ is given by
\begin{align}
	\Gamma_{k\rightarrow l}=\sum_{i}|\langle \gamma_l | i\rangle\langle i|\gamma_k\rangle|^{2}\Gamma_{i}(\Delta\epsilon_{kl})\ge 0.
\end{align}
Similarly, one can show that the dephasing rates $\Gamma_{\gamma_k,\gamma_l}$ of excitonic coherences $\ket{\gamma_k}\bra{\gamma_l}$ are expressed as
\begin{align}
	\Gamma_{\gamma_k,\gamma_l}&=\sum_{i}(|\langle i| \gamma_k\rangle |^{2}-|\langle i| \gamma_l\rangle |^{2})^{2}\Gamma_{i}+\frac{1}{2}\sum_{m\neq k}\Gamma_{k\rightarrow m}+\frac{1}{2}\sum_{m\neq l}\Gamma_{l\rightarrow m}\\
	&=\Gamma_{\gamma_k}+\Gamma_{\gamma_l}-2\sum_{i}|\langle \gamma_l | i\rangle\langle i|\gamma_k\rangle|^{2}\Gamma_i,
\end{align}
with $\Gamma_{\gamma_k,\gamma_l}\le \Gamma_{\gamma_k}+\Gamma_{\gamma_l}$, implying that the excitonic dephasing rate $\Gamma_{\gamma_k,\gamma_l}$ decreases as the spatial overlap between excitonic wavefunctions, $|\langle \gamma_l | i\rangle\langle i|\gamma_k\rangle|^{2}$, increases. These equations correspond to Eq.~(4) and (5) in the main text.

Based on the fact that there are notable spatial overlap between B850 excitons $\ket{\beta_k}$, while the spatial overlap between B800 and B850 excitons and that between B800 excitons $\ket{\alpha_k}$ are negligible, we employ a phenomenological Lindblad equation for the pure dephasing noise of the form
\begin{align}
	L_{pd}&=\sum_{\alpha_k}\Gamma_\alpha\left(A_{\alpha_k}\rho(t)A_{\alpha_k}^{\dagger}-\frac{1}{2}\{A_{\alpha_k}^{\dagger}A_{\alpha_k},\rho(t)\}\right)\\
	&\quad+\sum_{\beta_k}\Gamma_\beta(1-\eta)\left(A_{\beta_k}\rho(t)A_{\beta_k}^{\dagger}-\frac{1}{2}\{A_{\beta_k}^{\dagger}A_{\beta_k},\rho(t)\}\right)\nonumber\\
	&\quad+\Gamma_\beta\eta\left(A_{\eta}\rho(t)A_{\eta}^{\dagger}-\frac{1}{2}\{A_{\eta}^{\dagger}A_{\eta},\rho(t)\}\right),\nonumber\\
	A_{\alpha_k}&=\ket{\alpha_k}\bra{\alpha_k},\quad A_{\beta_k}=\ket{\beta_k}\bra{\beta_k},\quad A_{\eta}=\sum_{k}\ket{\beta_k}\bra{\beta_k},
\end{align}
where the dephasing rates $\Gamma_\alpha$ and $\Gamma_\beta$ of the B800 and B850 excitons are estimated based on absorption spectra (see Figure 1). This effective Lindblad model leads to the dephasing rates of $\Gamma_\alpha/2$ and $\Gamma_\beta/2$, respectively, for the optical coherences $\ket{g}\bra{\alpha_k}$ and $\ket{g}\bra{\beta_k}$, and the uncorrelated dephasing rates of $\Gamma_{\alpha}$ and $(\Gamma_{\alpha}+\Gamma_{\beta})/2$, respectively, for the excitonic coherences $\ket{\alpha_k}\bra{\alpha_{l\neq k}}$ between B800 excitons and $\ket{\alpha_k}\bra{\beta_l}$ between B800 and B850 excitons. For the excitonic coherences $\ket{\beta_k}\bra{\beta_{l\neq k}}$ between B850 excitons, the model leads to a correlated dephasing rate $\Gamma_{\beta}(1-\eta)$ where $\eta$ describes the degree of correlations in the noise due to the spatial overlap between excitonic wavefunctions, which is taken to be $1-\eta=\Gamma_{\beta_k,\beta_l}/(\Gamma_{\beta_k}+\Gamma_{\beta_l})\approx 0.7$ (see Appendix \ref{char_excs}). The relaxation of B850 excitons is also considered in the simulations with the relaxation rate of $20\,{\rm cm}^{-1}$, as discussed in Appendix \ref{char_excs}.

\subsubsection{The Huang-Rhys factor in the exciton basis}

In the main text, we consider vibronic Hamiltonian in the form
\begin{align}
	H_{e-v}&=\sum_{i}\sqrt{s}\omega(b_i+b_{i}^{\dagger})\ket{i}\bra{i}\\
	&=\sum_{k,l}\sqrt{S_{kl}}\omega(a_{kl}+a_{kl}^{\dagger})\ket{\gamma_{k}}\bra{\gamma_{l}},
\end{align}
where $\omega$ is the frequency of local vibrational modes and $s$ is the local Huang-Rhys factor that quantifies the vibronic coupling strength in the site basis. Using the orthogonality between exciton states, one can show that $\sqrt{S_{kl}}\omega(a_{kl}+a_{kl}^{\dagger})=\bra{\gamma_k}	H_{e-v}\ket{\gamma_l}=\sum_i \sqrt{s}\omega(b_i+b_{i}^{\dagger})\langle\gamma_k | i\rangle\langle i |\gamma_l\rangle$, leading to $a_{kl}=\sqrt{s/S_{kl}}\sum_{i}b_{i} \langle\gamma_k | i\rangle\langle i |\gamma_l\rangle$. Using the canonical commutation relations $[b_i,b_{j}^{\dagger}]=\delta_{i,j}$ with $\delta_{i,j}$ denoting the Kronecker delta ($\delta_{i,j}=1$ if $i=j$ and $\delta_{i,j}=0$ otherwise), one can show that 
\begin{align}
	[a_{kl},a_{kl}^{\dagger}]=\mathbb{1}=\frac{s}{S_{kl}}\sum_{i}|\langle\gamma_k | i\rangle\langle i |\gamma_l\rangle|^{2}\mathbb{1},
\end{align}
with $\mathbb{1}$ denoting an identity operator, where the generalized HR factor is reduced to $S_{kl}=s\sum_{i}|\langle\gamma_k | i\rangle\langle i |\gamma_l\rangle|^{2}$, as shown in the main text.

\section{Experiment characterization}

\subsection{Light-matter interaction. Experimental laser pulses}

The $Q_y$ transition dipoles $\vec{d}_i$ of individual pigments can be excited by a nearly resonant electric field $\vec{E}(t)$. The excitation process is described by the light 
matter interaction Hamiltonian 
\begin{equation}
H_{field}(t)=\vec E(t)\cdot(\sum_{k}\vec{D}_{\alpha_k}\ket{\alpha_k}\bra{g}+\sum_{\beta_k}\vec{D}_{\beta_k}\ket{\beta_k}\bra{g})+h.c,
\end{equation}
where $\vec{E}(t)=E(t)\hat{e}$ denote the electric field of the laser pulse with amplitude E(t) and polarization $\hat{e}$. The optical response of the system depends, according to $H_{field}$, on 
the transition 
dipoles $\vec{D}_{\beta_k}=\sum_{i\epsilon B850}^{N850} \bracket{\beta_k}{i}\ket{i}$ and $\vec{D}_{\alpha_k}=\sum_{i\epsilon B800} \bracket{\alpha_k}{i}\ket{i}$, whose specific values for the 
LH2 were already 
provided in Appendix \ref{char_excs}.

\begin{figure}
\centering
\begin{minipage}{1\columnwidth}
\includegraphics[width=.9\columnwidth]{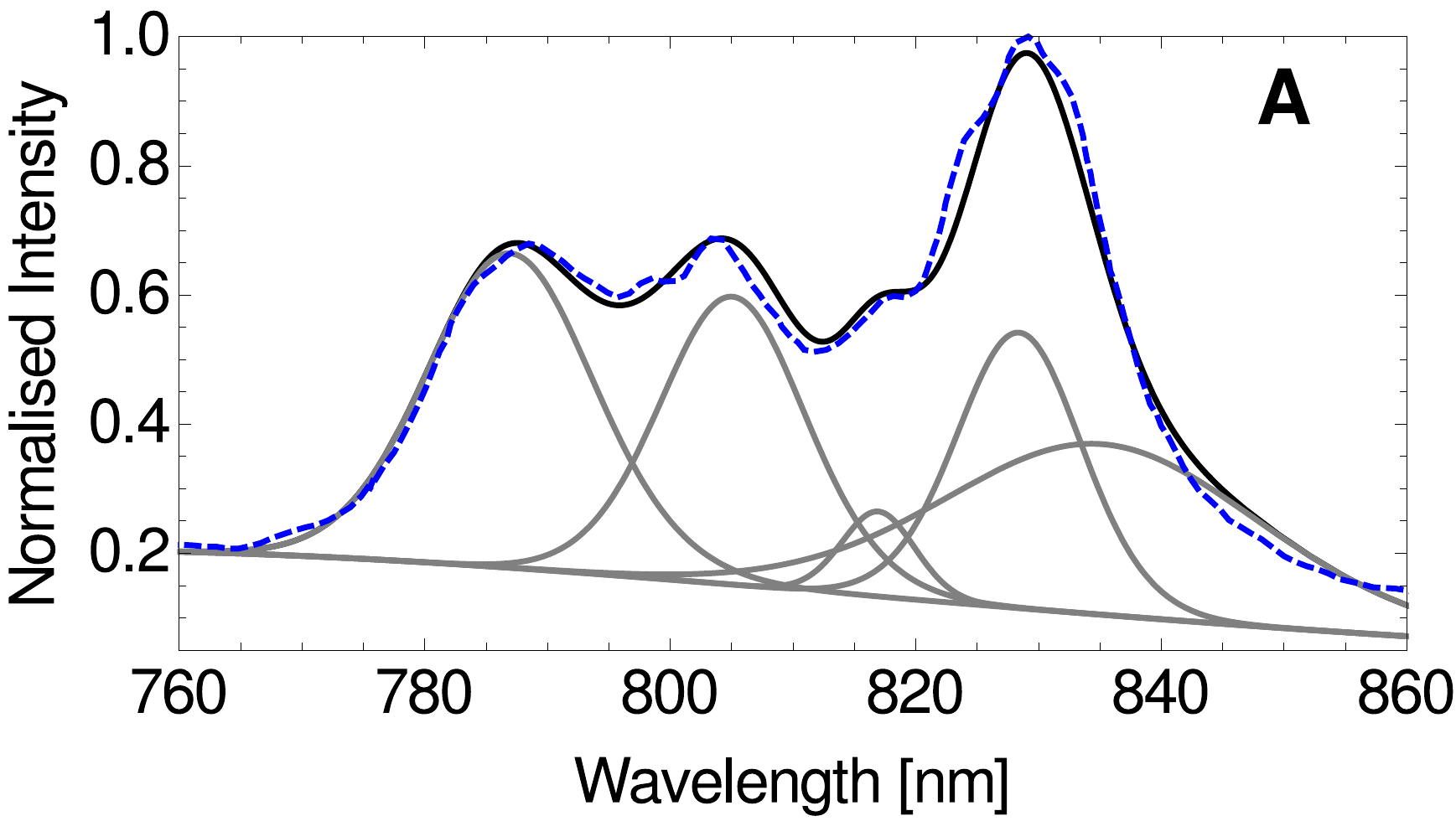}
\end{minipage}
\begin{minipage}{1\columnwidth}
\vspace{0.3 cm}
\includegraphics[width=1\columnwidth]{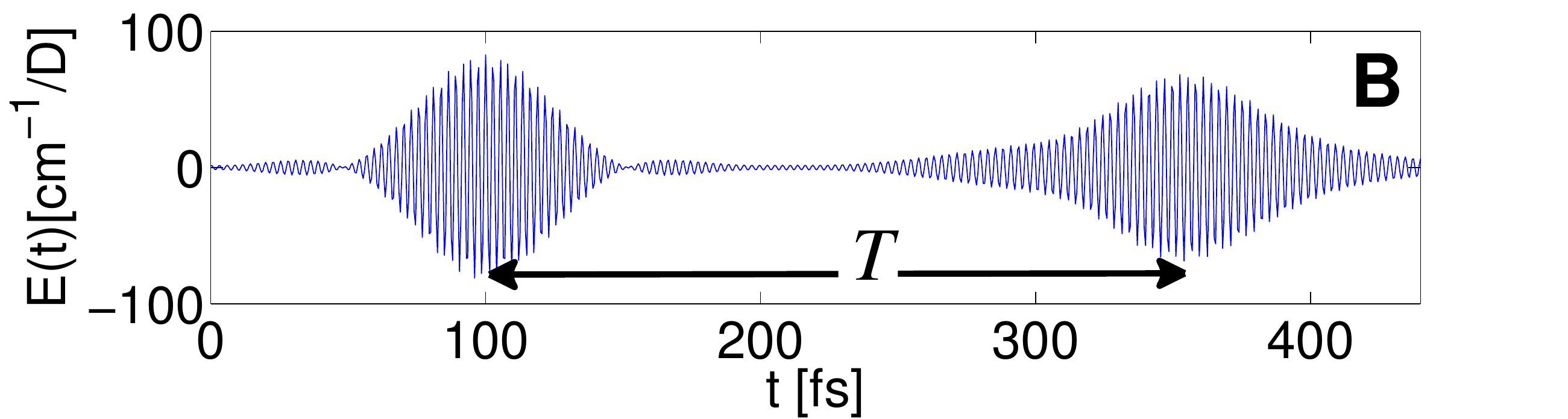}
\end{minipage}
\caption{Laser spectral profile. In {\bf A} the laser employed (dashed-blue) is presented along the fit (black) of the square modulus of the sum six Gaussians 
$E_i(\omega)=A_i\exp[-\tau^2(\omega-\omega_i)^2/2]$, with parameters $\{A_1,A_2,A_3,A_4,A_5,A_6\}=\{0.394,0.384,0.397,0.284,0.452,0.15\}$,  
$2\pi/\{\omega_1,\omega_2,\omega_3,\omega_4,\omega_5,\omega_6\}=\{787,805,828,836,750,817\}$ nm, and $\{\tau_1,\tau_2,\tau_3,\tau_4,\tau_5,\tau_6\}=\{42.73,51.04,62.31,25.56,3.20,111.81\}$ fs. Since 
the component $E_5(\omega)$ has a very narrow time width $\tau_5$, it becomes in the frequency domain, an offset for the other more structured components.  The addition of the offset and each electric 
field component squared  $|E_i(\omega)+E_5(\omega)|^2$, is presented in grey continuous lines. In {\bf B} the outcome of the phase ramp applied to these pulses with $\omega_K=2\pi/820$ nm, delays by a 
time $T$ (in this example $T=250$ fs) spectral components with longer wavelength than 820 nm$=2\pi/\omega_K$. The amplitude of the fields, given in cm$^{-1}$/Debye correspond to the laser intensity of 
200 W/cm$^2$ with $10^{-6}$ duty cycle \cite{Hildner2013}.  }
\label{fig2A}
\end{figure}

To correctly simulate the experimental excitation of the LH2 complexes as was done in \cite{Hildner2013}, illumination of the system has to be accounted for in full detail. The experiment features 
two distinct laser pulses intended to address the two different spectral regions composing the LH2; namely, the peaks at 850 and 800 nm. The two pulses where obtained from a single lock-in pulse 
covering the 760-860 spectral region \cite{Hildner2013}, with an spectral profile fitted  --as shown in Fig.\ref{fig2A}{\bf A}-- by the square modulus of a sum of electric fields $E(\omega)=\sum_i 
E_i(\omega)$, each modeled by a Gaussian function. The outcome pulses in the time domain are the result of Fourier transforming these Gaussians with or without an additional phase (which is a 
linear function of the frequency) depending on whether the carrier field frequency is smaller or greater than a kink frequency $\omega_K$; i.e. $E(t)=\int_{-\infty}^{\omega_K}d\omega \sum_i 
E(\omega)\exp[i\omega t+i(\phi - \omega s)]+\int^{\infty}_{\omega_K}d\omega \sum_i E(\omega)\exp[i\omega t ]$. The exponential on the first integral can thus be written as $\exp[i\omega t - s) + 
i\phi]$, where s introduces the time delay that separates the initial laser pulse into two different components acting on either side of the kink frequency $\omega_K$, and $\phi$ is a relative phase 
between pulses. Thus, for the two color 
experiment, the kink frequency was chosen at $2\pi/\omega_K=820$nm, generating a first pulse whose frequency spectrum was approximately flat in the 780 - 820 nm spectral region; and a second, 
time-delayed pulse with a frequency spectrum between 820 and 860 nm peaking at $\approx$ 828 nm. The resultant electric fields after this operation are 
presented for an specific time delay $T$ ($T\propto r$) in Fig.\ref{fig2A}{\bf B}. The scheme described was intended for producing two distinct laser pulses covering different spectral regions. 

\subsection{On the fast oscillating component in $FI(T)$ and the origin of the sinusoidal modulation of fluorescence intensity as a function of the pulse's relative phase $\phi$}

\begin{center}
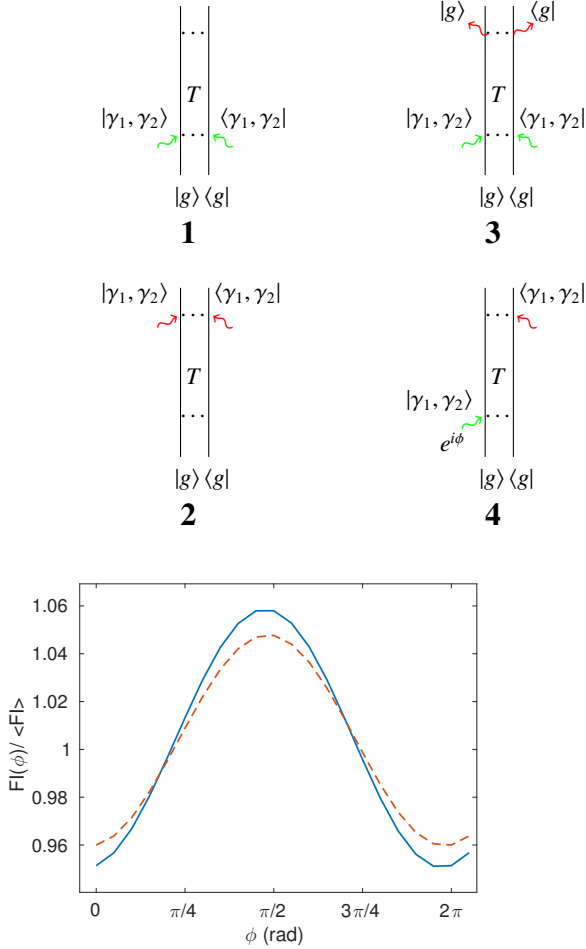
\begin{figure}
\begin{minipage}{.4\columnwidth}
\setlength{\unitlength}{0.75mm}
\begin{picture}(85,100)(7,0)
\put(14,62){${\ket{g}\bra{g}}$}
\put(15,67){\line(0,1){30}}
\put(20,67){\line(0,1){30}}
\put(15,74){\ldots}
\put(16,80){$T$}
\put(15,92){\ldots}
\put(11.5,71){\begin{rotate}{30}$\color{green}\huge\leadsto$\end{rotate}}
\put(24.8,73){\begin{rotate}{150}$\color{green}\leadsto$\end{rotate}}
\put(1,76){${\ket{\gamma_1,\gamma_2}}$}
\put(22,76){${\bra{\gamma_1,\gamma_2}}$}
\put(15,55){\large\bf 1}
\put(14,12){${\ket{g}\bra{g}}$}
\put(15,17){\line(0,1){30}}
\put(20,17){\line(0,1){30}}
\put(15,24){\ldots}
\put(16,30){$T$}
\put(15,42){\ldots}
\put(11.5,39){\begin{rotate}{30}$\color{red}\huge\leadsto$\end{rotate}}
\put(24.8,41){\begin{rotate}{150}$\color{red}\leadsto$\end{rotate}}
\put(1,45){${\ket{\gamma_1,\gamma_2}}$}
\put(21,45){${\bra{\gamma_1,\gamma_2}}$}
\put(15,5){\large\bf 2}

\put(68,62){${\ket{g}\bra{g}}$}
\put(69,67){\line(0,1){30}}
\put(74,67){\line(0,1){30}}
\put(69,74){\ldots}
\put(70,80){$T$}
\put(69,92){\ldots}
\put(70,93){\begin{rotate}{150}$\color{red}\leadsto$\end{rotate}}
\put(74.7,91){\begin{rotate}{30}$\color{red}\leadsto$\end{rotate}}
\put(65.5,71){\begin{rotate}{30}$\color{green}\leadsto$\end{rotate}}
\put(78.8,73){\begin{rotate}{150}$\color{green}\leadsto$\end{rotate}}
\put(55,76){$\ket{\gamma_1,\gamma_2}$}
\put(75,76){$\bra{\gamma_1,\gamma_2}$}
\put(61,95){$\ket{g}$}
\put(77,95){$\bra{g}$}

\put(69,55){\large\bf 3}

\put(68,12){${\ket{g}\bra{g}}$}
\put(69,17){\line(0,1){30}}
\put(74,17){\line(0,1){30}}
\put(69,24){\ldots}
\put(70,30){$T$}
\put(69,42){\ldots}
\put(78.7,41){\begin{rotate}{150}$\color{red}\leadsto$\end{rotate}}
\put(65.5,21){\begin{rotate}{30}$\color{green}\leadsto$\end{rotate}}
\put(55,26){$\ket{\gamma_1,\gamma_2}$}
\put(75,45){$\bra{\gamma_1,\gamma_2}$}
\put(61,18){$ e^{i\phi}$}
\put(69,5){\large\bf 4}
\end{picture}

\end{minipage}
\vspace{0.5 cm}
\includegraphics[width=.8\columnwidth]{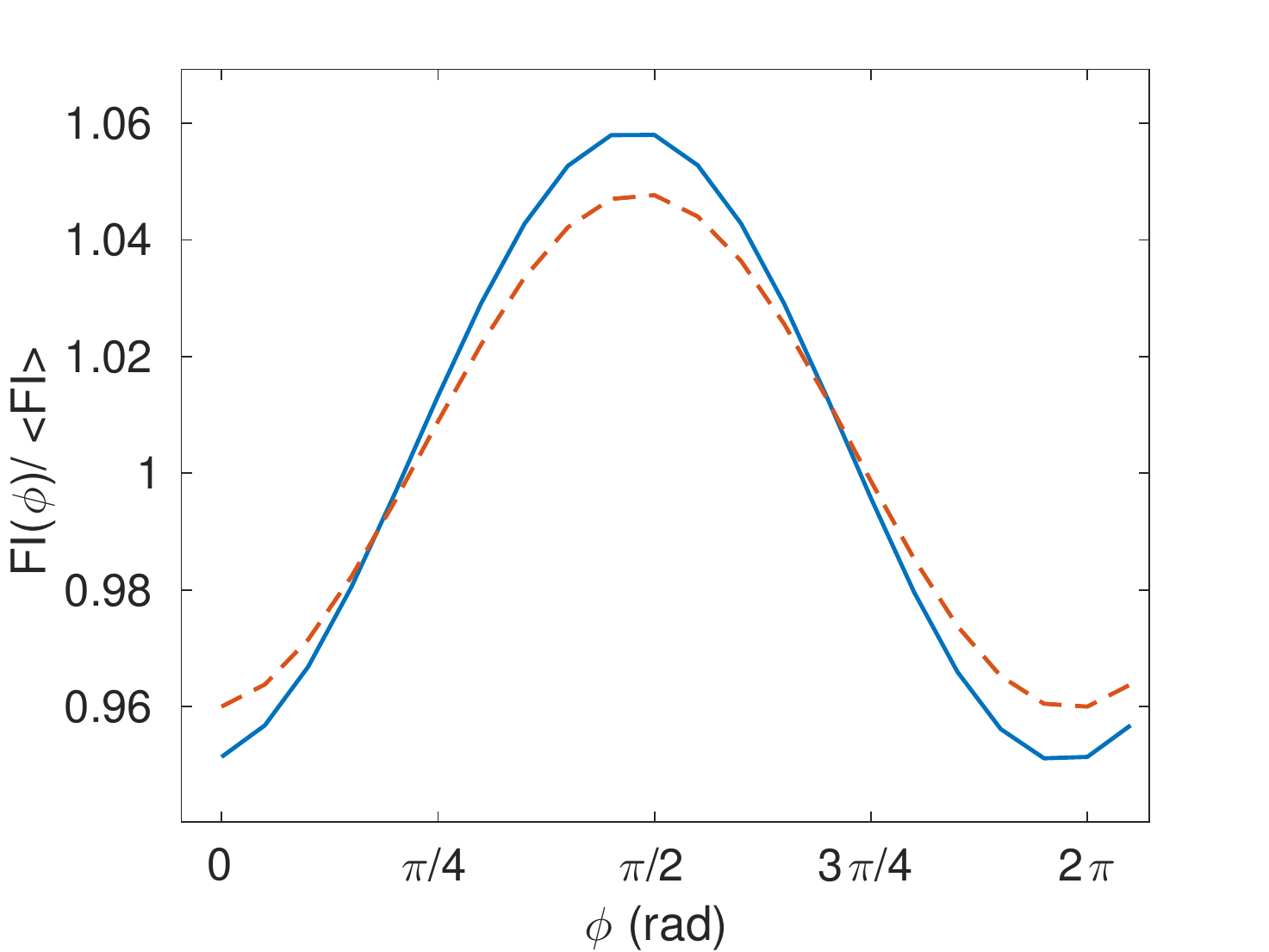}
\caption{The origin of relative pulses phase modulation of fluorescence intensity. The relevant Feynman diagrams for the two colour experiment present the interaction with the field (first and 
second pulses in green and red wiggly arrows, with relative phase $\phi$) useful for the description of oscillatory FI traces. Here, dotted lines  illustrate the times when transitions happen between 
ground $\ket{g}$ and excited kets $\ket{\gamma_{1}}$ or $\ket{\gamma_{2}}$, or respective bras. These times bound the inter-pulse time $T$ during which coherent population transfer 
$\ket{\gamma_1}\bra{\gamma_1}\leftrightarrow \ket{\gamma_2}\bra{\gamma_2}$  occurs. In the lower panel are presented the fluorescence intensity traces $FI(\phi)$ for the vibronic (continuous) and 
excitonic (dashed) models, regarding the pair of B850 excitons lying at 845-894 nm. This Figure is analogous to the inset of Fig. 3{\bf D} in the main text, but obtained with an electric field 
which is one fourth of that presented in Fig.\ref{fig2A}{\bf B}.}\label{fig3A}
\end{figure}
\end{center}

Simulations of fluorescence intensity for the two color experiment in \cite{Hildner2013} display --as shown in Fig. 3{\bf D} on the main text-- the expected frequency of $\approx$ 200 fs, 
together with a transient, fast optical frequency oscillating component at $\approx$ 3 fs, which at 250 fs is already much damped and which does not show in the experiments in \cite{Hildner2013} due 
to a longer sampling time. To understand the origin of this fast frequency component, it is illustrative to review the Feynman diagrams for the optical response of this system to the laser pulses 
described in the previous section. 

Feynman diagrams describe the complete evolution of the system, including interaction with the laser pulses and free dynamics, as shown in Fig.\ref{fig3A}{\bf A}. Rigorously,  these diagrams are 
devised  for the impulsive limit --i.e. for pulses with much shorter temporal duration than the time-scale of dynamics for the free Hamiltonian \cite{Mukamel}-- which is not the case here; 
nonetheless they serve to illustrate the multiple dynamics encoded in the $FI(T)$ traces. In the Feynman diagrams in Fig. \ref{fig3A}{\bf A} time evolves upwards, dotted lines correspond to the times 
at which impulsive excitation occurs (hence confining the inter-pulse delay time T of free dynamics), and wiggly lines correspond to  interactions with the electromagnetic field at 
either bras or kets of the evolving density operator. In Fig.\ref{fig3A}{\bf A}, green and red wiggly lines capture the interaction with the first or second pulses, which induce transitions between 
the ground 
$\ket{g}$ and the states $\ket{\gamma}$, with mirror transitions operating in the respective bras, i.e, $\bra{g},\bra{\gamma}$. After excitation, free Hamiltonian evolution ensues, with either bra 
$\bra{\gamma_1}\leftrightarrow\bra{\gamma_2}$ or ket $\ket{\gamma_1}\leftrightarrow\ket{\gamma_2}$ undergoing vibronic exchange of probability amplitudes.
The total excited state population, proportional to $FI(T,\phi)$, is obtained as a projection of the density operator after finishing the pulsed sequence, and only captures its diagonal elements. 
Simultaneous interactions of bra and ket with the first  pulse or with the second pulse, prompt populations or coherences in the excited state manifold, as shown by the diagrams {\bf 1} and {\bf 2}, 
respectively. The dynamics driven by a mode during $T$ will be read out whenever the second pulse produces an inversion, as shown in {\bf 3}. The contrast of {\bf 1} and {\bf 3} gives rise to the 
oscillations from populations in the FI traces.

Diagram {\bf 4} hints, however, towards dynamics during $T$ that do not occur exclusively in the excited state manifold. This diagram differs from the others in that only the ket (and not the bra) is 
excited with the first pulse, hence initializing a density operator contribution that evolves during $T$, instead of vibronically, as an excited-ground state coherence $\ket{g}\bra{\gamma}$, 
which after the second pulse is mapped into the excited state populations $\ket{\gamma}\bra{\gamma}$. Since the excited-ground state coherence oscillates 
at the optical frequencies, it is 
the origin of the fast oscillating but short lived component present in the traces of Figs. 3{\bf D} in the main text. Such fast oscillating contributions hence decay with an inverse rate of 120 fs, 
associated here to $\Gamma_\alpha/2=48$ or $(\Gamma_\beta+\Gamma_{relax})/2=49$ cm$^{-1}$ for B800  of B850 excitons, respectively.  Notice that this contribution is, as sketched in Diagram {\bf 4}, 
the only one which can encode the phase $\phi$, since it depends on the excited state amplitudes and not on their population. Since it is a relative phase 
between ground and excited states, it cannot be encoded in the excited state coherences, such as those created in diagrams {\bf 1}-{\bf 3}. The effect of this phase it to add a factor $e^{i\phi}$ to 
all excited state amplitudes, thus creating a phase difference between the excited and ground state amplitudes. 

The experimentally unresolved fast oscillating contribution in $FI(T)$ clarified above, is the only one that may generate the observed modulation of the fluorescence from variation of $\phi$. Higher 
order matter-field interactions may also contribute, but since the laser intensity used was selected below saturation, it is safe to assume that only the  lowest order contributions will be relevant. 
The contribution of diagram {\bf 4} outweighs that of diagram {\bf 3} for lower laser intensities, since it depends on two light-matter interactions, while diagram  {\bf 3} depends on four (note the 
wiggly lines in Fig.\ref{fig3A}{\bf A}), the consequence being that $FI(\phi)$ gains amplitude with lower light intensities. 

Figure \ref{fig3A}{\bf B} shows the phase-dependent dynamics $FI(\phi)$  at a fixed delay time, for the 844-795 nm pair of states; thus analogous to the inset shown in Fig. 3{\bf D} on the main 
text. Here however, the intensity of the simulated electric fields is one fourth of the estimated electric fields intensity in the experiments, which was used in all previous simulations (see 
Fig. \ref{fig2A}). Figure \ref{fig3A}{\bf B} therefore demonstrates the dependence of the FI on the intensity of the electric fields, as a reduction on the latter leads to an increase of two to three 
times for the amplitude of the $FI(\phi)$, as compared to the experimental electric fields intensity case. However, for the same reduced electric field intensity, the $FI(T)$ traces no longer shows 
clearly the 200 fs period oscillations, since diagram {\bf 3}, responsible for this long-lasting traces, reduces its amplitude proportionally to the laser intensity. 

The complex interplay between the $FI(T)$ and the $FI(\phi)$, together with their dependence on the laser intensity, pinpoints a limitation of our model, as reproducibility of the experimental 
results regarding the oscillation in $FI(T)$ leads to an estimation of the laser intensity which, if correct, would in turn mean that $FI(\phi)$ is not correctly reproduced (too small amplitude), 
while on the other hand, simulating a lower laser intensity leads to a correct amplitude for $FI(\phi)$ but no oscillations in $FI(T)$. As the experimental laser intensity is known and fixed, it is 
obvious that our model does not correctly capture all the components in $FI(\phi)$. Using pairs of excitons and a single mode coordinate is too simplistic, and a model including more degrees of 
freedom should increase the accuracy in reproducing $FI(\phi)$ without changing the results in $FI(T)$. In particular, incorporating B800 excitons, which present a similar dephasing as the B850 
excitons, will likely  increase the $FI(\phi)$ amplitude without changing $FI(T)$, as these excitons span a much narrower energy window with smaller energy gaps which are i) off-resonant with 
any discrete vibrations, which are usually of higher energy and ii) not resolved by the 
100-200 fs laser pulses used. 

\section{Control experiments}

\begin{figure}[!t]
\includegraphics[width=.7\columnwidth]{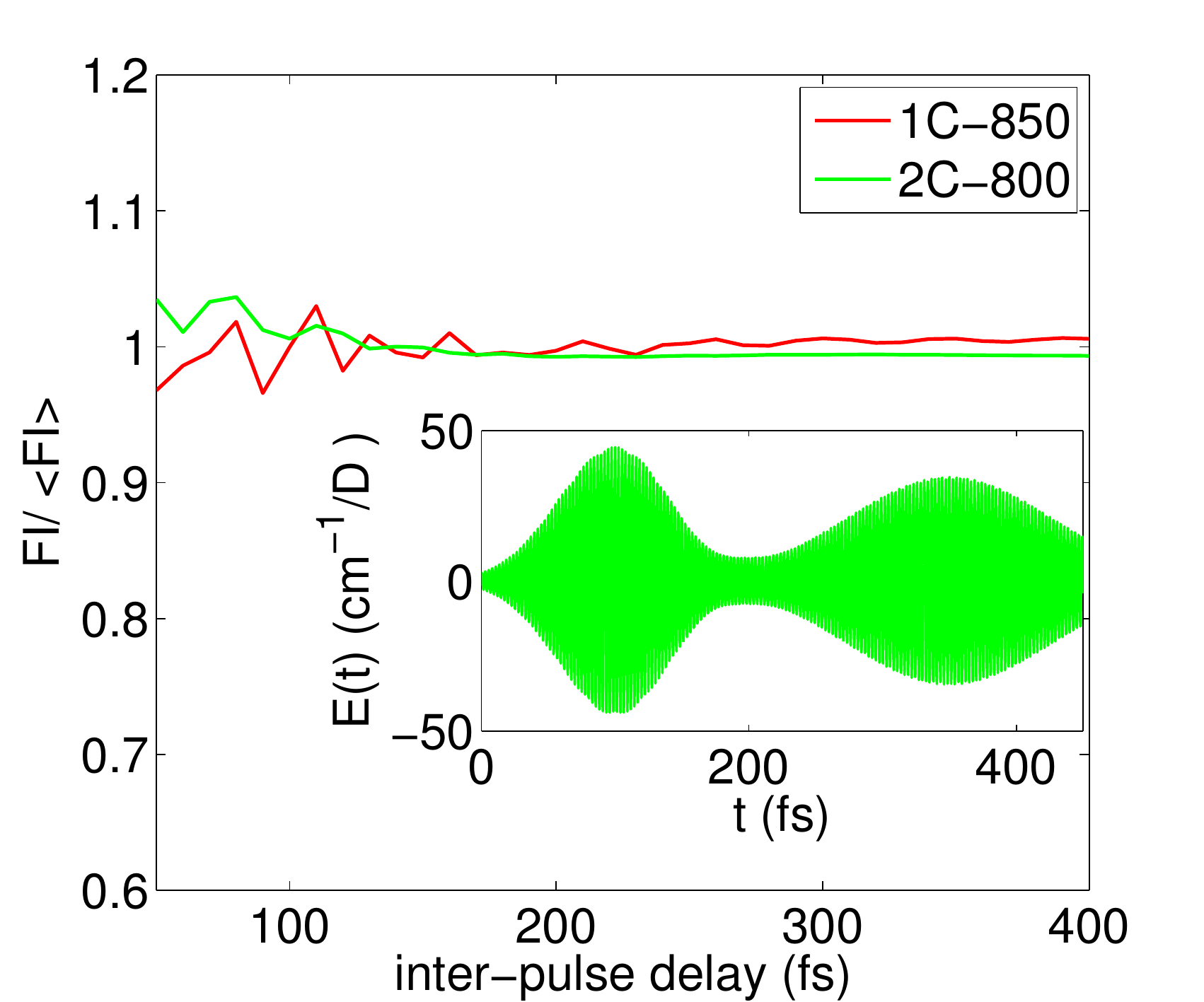}
\caption{Resultant fluorescence intensity traces as a function of pulses time delay, $FI(T)$, for the two colour 800 nm (2C-800) and one colour 850 nm (1C-850) experimental set-ups. 
These traces are the result from the addition  of the two  relevant vibronic contributions (845-794 nm and 838-788 nm) discussed  in Fig. 3{\bf C} in the main text. Inset shows the simulated laser 
pulses used in the 2C-800 experiments.}\label{fig4A}
\end{figure}

In \cite{Hildner2013} two control experiments were performed. In these experiments FI delay traces were obtained with two identical pulses (hence one colour) of wavelengths longer than 825 nm (1C-850), or two 
different pulses, one with an slightly shorter and another with an slightly longer wavelength than 800 nm (2C-800). No oscillation in the $FI(T)$ traces was observed for these control experiments.

To understand why no oscillations are observed whenever both laser pulses have similar wavelength, recall that the $FI(T)$ depends on the vibronic  interaction, and that the relevant modes have 
frequencies between 700-800 cm$^{-1}$. These conditions impose the restriction that the side-band must be excited in order to initiate the coherent exchange that leads to oscillations. Moreover, 
inversion to the ground state between side-bands, which has a dipole strength almost equal to the zero-phonon line transition, must be resonant with the read out second pulse. As a consequence, 
oscillatory dynamics is observable whenever pulses are spectrally separated by about the mode's frequency. Both control experimental schemes fail to simultaneously excite the side-band (possible with 
any of the $\simeq$ 800 nm pulses) and read out the subsequent coherent population exchange (possible with the $\simeq$ 830 nm  pulse). Therefore, the control experiments do not result in observable 
oscillations indicating the coherent population exchange, as displayed in Fig.\ref{fig4A}.

To create the pulses that are used in the 2C-800 control experiments, \cite{Hildner2013} had to use a phase ramp scheme with a linear phase below the 800 nm, instead of the 820 described 
above, that leads to much longer pulses than the original first pulse in the main experiment, as shown in the inset of Fig.\ref{fig4A} (compare with the pulse sequence shown in Fig.\ref{fig2A}{\bf 
B}). Notice in particular that the second pulse extends up to a width of about 200 fs, hence is spectrally narrower and specific to the 800 nm range; thereby, unable to induce the population inversion 
between side-bands required to resolve the vibronic dynamics.

\bibliography{LH2_references}

\begin{thebibliography}{52}%
\makeatletter
\providecommand \@ifxundefined [1]{%
 \@ifx{#1\undefined}
}%
\providecommand \@ifnum [1]{%
 \ifnum #1\expandafter \@firstoftwo
 \else \expandafter \@secondoftwo
 \fi
}%
\providecommand \@ifx [1]{%
 \ifx #1\expandafter \@firstoftwo
 \else \expandafter \@secondoftwo
 \fi
}%
\providecommand \natexlab [1]{#1}%
\providecommand \enquote  [1]{``#1''}%
\providecommand \bibnamefont  [1]{#1}%
\providecommand \bibfnamefont [1]{#1}%
\providecommand \citenamefont [1]{#1}%
\providecommand \href@noop [0]{\@secondoftwo}%
\providecommand \href [0]{\begingroup \@sanitize@url \@href}%
\providecommand \@href[1]{\@@startlink{#1}\@@href}%
\providecommand \@@href[1]{\endgroup#1\@@endlink}%
\providecommand \@sanitize@url [0]{\catcode `\\12\catcode `\$12\catcode
  `\&12\catcode `\#12\catcode `\^12\catcode `\_12\catcode `\%12\relax}%
\providecommand \@@startlink[1]{}%
\providecommand \@@endlink[0]{}%
\providecommand \url  [0]{\begingroup\@sanitize@url \@url }%
\providecommand \@url [1]{\endgroup\@href {#1}{\urlprefix }}%
\providecommand \urlprefix  [0]{URL }%
\providecommand \Eprint [0]{\href }%
\providecommand \doibase [0]{http://dx.doi.org/}%
\providecommand \selectlanguage [0]{\@gobble}%
\providecommand \bibinfo  [0]{\@secondoftwo}%
\providecommand \bibfield  [0]{\@secondoftwo}%
\providecommand \translation [1]{[#1]}%
\providecommand \BibitemOpen [0]{}%
\providecommand \bibitemStop [0]{}%
\providecommand \bibitemNoStop [0]{.\EOS\space}%
\providecommand \EOS [0]{\spacefactor3000\relax}%
\providecommand \BibitemShut  [1]{\csname bibitem#1\endcsname}%
\let\auto@bib@innerbib\@empty
\bibitem [{\citenamefont {{Engel}}\ \emph {et~al.}(2007)\citenamefont
  {{Engel}}, \citenamefont {{ Calhoun}}, \citenamefont {{Read}}, \citenamefont
  {{Ahn}}, \citenamefont {{Man{\v c}al}}, \citenamefont {{Cheng}},
  \citenamefont {{Blankenship}},\ and\ \citenamefont
  {{Fleming}}}]{Engel_Nature2006}%
  \BibitemOpen
  \bibfield  {author} {\bibinfo {author} {\bibfnamefont {G.}~\bibnamefont
  {{Engel}}}, \bibinfo {author} {\bibfnamefont {T.}~\bibnamefont {{ Calhoun}}},
  \bibinfo {author} {\bibfnamefont {E.}~\bibnamefont {{Read}}}, \bibinfo
  {author} {\bibfnamefont {T.}~\bibnamefont {{Ahn}}}, \bibinfo {author}
  {\bibfnamefont {T.}~\bibnamefont {{Man{\v c}al}}}, \bibinfo {author}
  {\bibfnamefont {Y.}~\bibnamefont {{Cheng}}}, \bibinfo {author} {\bibfnamefont
  {R.}~\bibnamefont {{Blankenship}}}, \ and\ \bibinfo {author} {\bibfnamefont
  {G.}~\bibnamefont {{Fleming}}},\ }\href@noop {} {\bibfield  {journal}
  {\bibinfo  {journal} {Nature}\ }\textbf {\bibinfo {volume} {446}},\ \bibinfo
  {pages} {782} (\bibinfo {year} {2007})}\BibitemShut {NoStop}%
\bibitem [{\citenamefont {{Hayes}}\ \emph {et~al.}(2010)\citenamefont
  {{Hayes}}, \citenamefont {{Panitchayangkoon}}, \citenamefont {{Fransted}},
  \citenamefont {{Caram}}, \citenamefont {{Wen}}, \citenamefont {{Freed}},\
  and\ \citenamefont {{Engel}}}]{EngelNJP}%
  \BibitemOpen
  \bibfield  {author} {\bibinfo {author} {\bibfnamefont {D.}~\bibnamefont
  {{Hayes}}}, \bibinfo {author} {\bibfnamefont {G.}~\bibnamefont
  {{Panitchayangkoon}}}, \bibinfo {author} {\bibfnamefont {K.~A.}\ \bibnamefont
  {{Fransted}}}, \bibinfo {author} {\bibfnamefont {J.~R.}\ \bibnamefont
  {{Caram}}}, \bibinfo {author} {\bibfnamefont {J.}~\bibnamefont {{Wen}}},
  \bibinfo {author} {\bibfnamefont {K.~F.}\ \bibnamefont {{Freed}}}, \ and\
  \bibinfo {author} {\bibfnamefont {G.}~\bibnamefont {{Engel}}},\ }\href@noop
  {} {\bibfield  {journal} {\bibinfo  {journal} {New J. Phys.}\ }\textbf
  {\bibinfo {volume} {12}},\ \bibinfo {pages} {065042} (\bibinfo {year}
  {2010})}\BibitemShut {NoStop}%
\bibitem [{\citenamefont {{Brixner}}\ \emph {et~al.}(2005)\citenamefont
  {{Brixner}}, \citenamefont {{Stenger}}, \citenamefont {{Vaswani}},
  \citenamefont {{Cho}}, \citenamefont {{Blankenship}},\ and\ \citenamefont
  {{Fleming}}}]{Fleming2005}%
  \BibitemOpen
  \bibfield  {author} {\bibinfo {author} {\bibfnamefont {T.}~\bibnamefont
  {{Brixner}}}, \bibinfo {author} {\bibfnamefont {J.}~\bibnamefont
  {{Stenger}}}, \bibinfo {author} {\bibfnamefont {H.~M.}\ \bibnamefont
  {{Vaswani}}}, \bibinfo {author} {\bibfnamefont {M.}~\bibnamefont {{Cho}}},
  \bibinfo {author} {\bibfnamefont {R.~E.}\ \bibnamefont {{Blankenship}}}, \
  and\ \bibinfo {author} {\bibfnamefont {G.~R.}\ \bibnamefont {{Fleming}}},\
  }\href@noop {} {\bibfield  {journal} {\bibinfo  {journal} {Nature}\ }\textbf
  {\bibinfo {volume} {434}},\ \bibinfo {pages} {625} (\bibinfo {year}
  {2005})}\BibitemShut {NoStop}%
\bibitem [{\citenamefont {{Harel}}\ and\ \citenamefont
  {{Engel}}(2012)}]{Engel_PNAS2012}%
  \BibitemOpen
  \bibfield  {author} {\bibinfo {author} {\bibfnamefont {E.}~\bibnamefont
  {{Harel}}}\ and\ \bibinfo {author} {\bibfnamefont {G.~S.}\ \bibnamefont
  {{Engel}}},\ }\href@noop {} {\bibfield  {journal} {\bibinfo  {journal} {Proc.
  Natl Acad. Sci. USA}\ }\textbf {\bibinfo {volume} {109}},\ \bibinfo {pages}
  {706} (\bibinfo {year} {2012})}\BibitemShut {NoStop}%
\bibitem [{\citenamefont {{Lee}}\ \emph {et~al.}(2007)\citenamefont {{Lee}},
  \citenamefont {{Cheng}},\ and\ \citenamefont
  {{Flemming}}}]{Flemming_2007Science}%
  \BibitemOpen
  \bibfield  {author} {\bibinfo {author} {\bibfnamefont {H.}~\bibnamefont
  {{Lee}}}, \bibinfo {author} {\bibfnamefont {Y.~C.}\ \bibnamefont {{Cheng}}},
  \ and\ \bibinfo {author} {\bibfnamefont {G.~R.}\ \bibnamefont {{Flemming}}},\
  }\href@noop {} {\bibfield  {journal} {\bibinfo  {journal} {Science}\ }\textbf
  {\bibinfo {volume} {316}},\ \bibinfo {pages} {1462} (\bibinfo {year}
  {2007})}\BibitemShut {NoStop}%
\bibitem [{\citenamefont {{Fuller}}\ \emph {et~al.}(2014)\citenamefont
  {{Fuller}}, \citenamefont {{Pan}}, \citenamefont {{Gelzinis}}, \citenamefont
  {{Butkus}}, \citenamefont {{Senlik}}, \citenamefont {{Wilcox}}, \citenamefont
  {{Yocum}}, \citenamefont {{Valkunas}}, \citenamefont {{Abramavicius}},\ and\
  \citenamefont {{Ogilvie}}}]{Ogilvie_NChem2014}%
  \BibitemOpen
  \bibfield  {author} {\bibinfo {author} {\bibfnamefont {F.~D.}\ \bibnamefont
  {{Fuller}}}, \bibinfo {author} {\bibfnamefont {J.}~\bibnamefont {{Pan}}},
  \bibinfo {author} {\bibfnamefont {A.}~\bibnamefont {{Gelzinis}}}, \bibinfo
  {author} {\bibfnamefont {V.}~\bibnamefont {{Butkus}}}, \bibinfo {author}
  {\bibfnamefont {S.~S.}\ \bibnamefont {{Senlik}}}, \bibinfo {author}
  {\bibfnamefont {D.~E.}\ \bibnamefont {{Wilcox}}}, \bibinfo {author}
  {\bibfnamefont {C.~F.}\ \bibnamefont {{Yocum}}}, \bibinfo {author}
  {\bibfnamefont {L.}~\bibnamefont {{Valkunas}}}, \bibinfo {author}
  {\bibfnamefont {D.}~\bibnamefont {{Abramavicius}}}, \ and\ \bibinfo {author}
  {\bibfnamefont {J.~P.}\ \bibnamefont {{Ogilvie}}},\ }\href@noop {} {\bibfield
   {journal} {\bibinfo  {journal} {Nature Chem.}\ }\textbf {\bibinfo {volume}
  {6}},\ \bibinfo {pages} {706} (\bibinfo {year} {2014})}\BibitemShut {NoStop}%
\bibitem [{\citenamefont {{Romero}}\ \emph {et~al.}(2014)\citenamefont
  {{Romero}}, \citenamefont {{Augulis}}, \citenamefont {{Novoderezhkin}},
  \citenamefont {{Ferretti}}, \citenamefont {{Thieme}}, \citenamefont
  {{Zigmantas}},\ and\ \citenamefont {{van Grondelle}}}]{Romero_NPhys2014}%
  \BibitemOpen
  \bibfield  {author} {\bibinfo {author} {\bibfnamefont {E.}~\bibnamefont
  {{Romero}}}, \bibinfo {author} {\bibfnamefont {R.}~\bibnamefont {{Augulis}}},
  \bibinfo {author} {\bibfnamefont {V.~I.}\ \bibnamefont {{Novoderezhkin}}},
  \bibinfo {author} {\bibfnamefont {M.}~\bibnamefont {{Ferretti}}}, \bibinfo
  {author} {\bibfnamefont {J.}~\bibnamefont {{Thieme}}}, \bibinfo {author}
  {\bibfnamefont {D.}~\bibnamefont {{Zigmantas}}}, \ and\ \bibinfo {author}
  {\bibfnamefont {R.}~\bibnamefont {{van Grondelle}}},\ }\href@noop {}
  {\bibfield  {journal} {\bibinfo  {journal} {Nature Phys.}\ }\textbf {\bibinfo
  {volume} {10}},\ \bibinfo {pages} {676} (\bibinfo {year} {2014})}\BibitemShut
  {NoStop}%
\bibitem [{\citenamefont {Duan}\ \emph {et~al.}(2017)\citenamefont {Duan},
  \citenamefont {Prokhorenko}, \citenamefont {Cogdell}, \citenamefont {Ashraf},
  \citenamefont {Stevens}, \citenamefont {Thorwart},\ and\ \citenamefont
  {Miller}}]{MIller_PNAS2017}%
  \BibitemOpen
  \bibfield  {author} {\bibinfo {author} {\bibfnamefont {H.-G.}\ \bibnamefont
  {Duan}}, \bibinfo {author} {\bibfnamefont {V.~I.}\ \bibnamefont
  {Prokhorenko}}, \bibinfo {author} {\bibfnamefont {R.~J.}\ \bibnamefont
  {Cogdell}}, \bibinfo {author} {\bibfnamefont {K.}~\bibnamefont {Ashraf}},
  \bibinfo {author} {\bibfnamefont {A.~L.}\ \bibnamefont {Stevens}}, \bibinfo
  {author} {\bibfnamefont {M.}~\bibnamefont {Thorwart}}, \ and\ \bibinfo
  {author} {\bibfnamefont {R.~J.~D.}\ \bibnamefont {Miller}},\ }\href {\doibase
  10.1073/pnas.1702261114} {\bibfield  {journal} {\bibinfo  {journal}
  {Proceedings of the National Academy of Sciences}\ }\textbf {\bibinfo
  {volume} {114}},\ \bibinfo {pages} {8493} (\bibinfo {year} {2017})},\ \Eprint
  {http://arxiv.org/abs/http://www.pnas.org/content/114/32/8493.full.pdf}
  {http://www.pnas.org/content/114/32/8493.full.pdf} \BibitemShut {NoStop}%
\bibitem [{\citenamefont {{Thyrhaug}}\ \emph {et~al.}(2017)\citenamefont
  {{Thyrhaug}}, \citenamefont {{Tempelaar}}, \citenamefont {{Alcocer}},
  \citenamefont {{{\v Z}{\'{\i}}dek}}, \citenamefont {{B{\'{\i}}na}},
  \citenamefont {{Knoester}}, \citenamefont {{Jansen}},\ and\ \citenamefont
  {{Zigmantas}}}]{Donatas_2017}%
  \BibitemOpen
  \bibfield  {author} {\bibinfo {author} {\bibfnamefont {E.}~\bibnamefont
  {{Thyrhaug}}}, \bibinfo {author} {\bibfnamefont {R.}~\bibnamefont
  {{Tempelaar}}}, \bibinfo {author} {\bibfnamefont {M.}~\bibnamefont
  {{Alcocer}}}, \bibinfo {author} {\bibfnamefont {K.}~\bibnamefont {{{\v
  Z}{\'{\i}}dek}}}, \bibinfo {author} {\bibfnamefont {D.}~\bibnamefont
  {{B{\'{\i}}na}}}, \bibinfo {author} {\bibfnamefont {J.}~\bibnamefont
  {{Knoester}}}, \bibinfo {author} {\bibfnamefont {T.~L.~C.}\ \bibnamefont
  {{Jansen}}}, \ and\ \bibinfo {author} {\bibfnamefont {D.}~\bibnamefont
  {{Zigmantas}}},\ }\href@noop {} {\bibfield  {journal} {\bibinfo  {journal}
  {arXiv e-prints}\ ,\ \bibinfo {pages}
  {http://adsabs.harvard.edu/abs/2017arXiv170900318Tt (Accessed 29 Nov 2017)}}
  (\bibinfo {year} {2017})},\ \Eprint {http://arxiv.org/abs/1709.00318}
  {arXiv:1709.00318 [physics.chem-ph]} \BibitemShut {NoStop}%
\bibitem [{\citenamefont {{Ishizaki}}\ \emph {et~al.}(2010)\citenamefont
  {{Ishizaki}}, \citenamefont {{Calhoun}}, \citenamefont {{Schlau-Cohen}},\
  and\ \citenamefont {{Fleming}}}]{Ishizaki2010}%
  \BibitemOpen
  \bibfield  {author} {\bibinfo {author} {\bibfnamefont {A.}~\bibnamefont
  {{Ishizaki}}}, \bibinfo {author} {\bibfnamefont {T.~R.}\ \bibnamefont
  {{Calhoun}}}, \bibinfo {author} {\bibfnamefont {G.~S.}\ \bibnamefont
  {{Schlau-Cohen}}}, \ and\ \bibinfo {author} {\bibfnamefont {G.~R.}\
  \bibnamefont {{Fleming}}},\ }\href@noop {} {\bibfield  {journal} {\bibinfo
  {journal} {Phys. Chem. Chem. Phys.}\ }\textbf {\bibinfo {volume} {12}},\
  \bibinfo {pages} {7319} (\bibinfo {year} {2010})}\BibitemShut {NoStop}%
\bibitem [{\citenamefont {{Chin}}\ \emph {et~al.}(2013)\citenamefont {{Chin}},
  \citenamefont {{Prior}}, \citenamefont {{Rosenbach}}, \citenamefont
  {{Caycedo-Soler}}, \citenamefont {{Huelga}},\ and\ \citenamefont
  {{Plenio}}}]{PlenioNature2013}%
  \BibitemOpen
  \bibfield  {author} {\bibinfo {author} {\bibfnamefont {A.~W.}\ \bibnamefont
  {{Chin}}}, \bibinfo {author} {\bibfnamefont {P.}~\bibnamefont {{Prior}}},
  \bibinfo {author} {\bibfnamefont {R.}~\bibnamefont {{Rosenbach}}}, \bibinfo
  {author} {\bibfnamefont {F.}~\bibnamefont {{Caycedo-Soler}}}, \bibinfo
  {author} {\bibfnamefont {S.~F.}\ \bibnamefont {{Huelga}}}, \ and\ \bibinfo
  {author} {\bibfnamefont {M.~B.}\ \bibnamefont {{Plenio}}},\ }\href@noop {}
  {\bibfield  {journal} {\bibinfo  {journal} {Nature Physics}\ }\textbf
  {\bibinfo {volume} {9}},\ \bibinfo {pages} {113} (\bibinfo {year}
  {2013})}\BibitemShut {NoStop}%
\bibitem [{\citenamefont {{Christensson}}\ \emph {et~al.}(2012)\citenamefont
  {{Christensson}}, \citenamefont {{Kauffmann}}, \citenamefont {{Pullerits}},\
  and\ \citenamefont {{Man{\v c}al}}}]{Christensson_JPCB2012}%
  \BibitemOpen
  \bibfield  {author} {\bibinfo {author} {\bibfnamefont {N.}~\bibnamefont
  {{Christensson}}}, \bibinfo {author} {\bibfnamefont {H.~F.}\ \bibnamefont
  {{Kauffmann}}}, \bibinfo {author} {\bibfnamefont {T.}~\bibnamefont
  {{Pullerits}}}, \ and\ \bibinfo {author} {\bibfnamefont {T.}~\bibnamefont
  {{Man{\v c}al}}},\ }\href@noop {} {\bibfield  {journal} {\bibinfo  {journal}
  {J. Phys. Chem. B}\ }\textbf {\bibinfo {volume} {116}},\ \bibinfo {pages}
  {7449} (\bibinfo {year} {2012})}\BibitemShut {NoStop}%
\bibitem [{\citenamefont {{Chenu}}\ \emph {et~al.}(2013)\citenamefont
  {{Chenu}}, \citenamefont {{Christensson}}, \citenamefont {{Kauffmann}},\ and\
  \citenamefont {{Man{\v c}al}}}]{Chenu_2013SciRep}%
  \BibitemOpen
  \bibfield  {author} {\bibinfo {author} {\bibfnamefont {A.}~\bibnamefont
  {{Chenu}}}, \bibinfo {author} {\bibfnamefont {N.}~\bibnamefont
  {{Christensson}}}, \bibinfo {author} {\bibfnamefont {H.}~\bibnamefont
  {{Kauffmann}}}, \ and\ \bibinfo {author} {\bibfnamefont {T.}~\bibnamefont
  {{Man{\v c}al}}},\ }\href@noop {} {\bibfield  {journal} {\bibinfo  {journal}
  {Sci. Rep.}\ }\textbf {\bibinfo {volume} {3}},\ \bibinfo {pages} {2029}
  (\bibinfo {year} {2013})}\BibitemShut {NoStop}%
\bibitem [{\citenamefont {{Plenio}}\ \emph {et~al.}(2013)\citenamefont
  {{Plenio}}, \citenamefont {{Almeida}},\ and\ \citenamefont
  {{Huelga}}}]{Plenio_2013JCP}%
  \BibitemOpen
  \bibfield  {author} {\bibinfo {author} {\bibfnamefont {M.~B.}\ \bibnamefont
  {{Plenio}}}, \bibinfo {author} {\bibfnamefont {J.}~\bibnamefont {{Almeida}}},
  \ and\ \bibinfo {author} {\bibfnamefont {S.~F.}\ \bibnamefont {{Huelga}}},\
  }\href@noop {} {\bibfield  {journal} {\bibinfo  {journal} {J. Chem. Phys.}\
  }\textbf {\bibinfo {volume} {139}},\ \bibinfo {pages} {235102} (\bibinfo
  {year} {2013})}\BibitemShut {NoStop}%
\bibitem [{\citenamefont {{Tiwari}}\ \emph {et~al.}(2013)\citenamefont
  {{Tiwari}}, \citenamefont {{Peters}},\ and\ \citenamefont
  {{Jonas}}}]{Tiwari2013}%
  \BibitemOpen
  \bibfield  {author} {\bibinfo {author} {\bibfnamefont {V.}~\bibnamefont
  {{Tiwari}}}, \bibinfo {author} {\bibfnamefont {W.~K.}\ \bibnamefont
  {{Peters}}}, \ and\ \bibinfo {author} {\bibfnamefont {D.~M.}\ \bibnamefont
  {{Jonas}}},\ }\href@noop {} {\bibfield  {journal} {\bibinfo  {journal}
  {Proceedings of the National Academy of Sciences}\ }\textbf {\bibinfo
  {volume} {110}},\ \bibinfo {pages} {1203} (\bibinfo {year}
  {2013})}\BibitemShut {NoStop}%
\bibitem [{\citenamefont {Novelli}\ \emph {et~al.}(2015)\citenamefont
  {Novelli}, \citenamefont {Nazir}, \citenamefont {Richards}, \citenamefont
  {Roozbeh}, \citenamefont {Wilk}, \citenamefont {Curmi},\ and\ \citenamefont
  {Davis}}]{Novelli_JPCLett2015}%
  \BibitemOpen
  \bibfield  {author} {\bibinfo {author} {\bibfnamefont {F.}~\bibnamefont
  {Novelli}}, \bibinfo {author} {\bibfnamefont {A.}~\bibnamefont {Nazir}},
  \bibinfo {author} {\bibfnamefont {G.~H.}\ \bibnamefont {Richards}}, \bibinfo
  {author} {\bibfnamefont {A.}~\bibnamefont {Roozbeh}}, \bibinfo {author}
  {\bibfnamefont {K.~E.}\ \bibnamefont {Wilk}}, \bibinfo {author}
  {\bibfnamefont {P.~M.~G.}\ \bibnamefont {Curmi}}, \ and\ \bibinfo {author}
  {\bibfnamefont {J.~A.}\ \bibnamefont {Davis}},\ }\href {\doibase
  10.1021/acs.jpclett.5b02058} {\bibfield  {journal} {\bibinfo  {journal} {The
  Journal of Physical Chemistry Letters}\ }\textbf {\bibinfo {volume} {6}},\
  \bibinfo {pages} {4573} (\bibinfo {year} {2015})},\ \bibinfo {note} {pMID:
  26528956},\ \Eprint
  {http://arxiv.org/abs/http://dx.doi.org/10.1021/acs.jpclett.5b02058}
  {http://dx.doi.org/10.1021/acs.jpclett.5b02058} \BibitemShut {NoStop}%
\bibitem [{\citenamefont {Prior}\ \emph {et~al.}(2010)\citenamefont {Prior},
  \citenamefont {Chin}, \citenamefont {Huelga},\ and\ \citenamefont
  {Plenio}}]{Prior}%
  \BibitemOpen
  \bibfield  {author} {\bibinfo {author} {\bibfnamefont {J.}~\bibnamefont
  {Prior}}, \bibinfo {author} {\bibfnamefont {A.~W.}\ \bibnamefont {Chin}},
  \bibinfo {author} {\bibfnamefont {S.~F.}\ \bibnamefont {Huelga}}, \ and\
  \bibinfo {author} {\bibfnamefont {M.~B.}\ \bibnamefont {Plenio}},\ }\href
  {\doibase 10.1103/PhysRevLett.105.050404} {\bibfield  {journal} {\bibinfo
  {journal} {Phys. Rev. Lett.}\ }\textbf {\bibinfo {volume} {105}},\ \bibinfo
  {pages} {050404} (\bibinfo {year} {2010})}\BibitemShut {NoStop}%
\bibitem [{\citenamefont {Chin}\ \emph {et~al.}(2010)\citenamefont {Chin},
  \citenamefont {Datta}, \citenamefont {Caruso}, \citenamefont {Huelga},\ and\
  \citenamefont {Plenio}}]{Chin}%
  \BibitemOpen
  \bibfield  {author} {\bibinfo {author} {\bibfnamefont {A.~W.}\ \bibnamefont
  {Chin}}, \bibinfo {author} {\bibfnamefont {A.}~\bibnamefont {Datta}},
  \bibinfo {author} {\bibfnamefont {F.}~\bibnamefont {Caruso}}, \bibinfo
  {author} {\bibfnamefont {S.~F.}\ \bibnamefont {Huelga}}, \ and\ \bibinfo
  {author} {\bibfnamefont {M.~B.}\ \bibnamefont {Plenio}},\ }\href
  {http://stacks.iop.org/1367-2630/12/i=6/a=065002} {\bibfield  {journal}
  {\bibinfo  {journal} {New Journal of Physics}\ }\textbf {\bibinfo {volume}
  {12}},\ \bibinfo {pages} {065002} (\bibinfo {year} {2010})}\BibitemShut
  {NoStop}%
\bibitem [{\citenamefont {{Huelga}}\ and\ \citenamefont
  {{Plenio}}(2013)}]{HuelgaP13}%
  \BibitemOpen
  \bibfield  {author} {\bibinfo {author} {\bibfnamefont {S.~F.}\ \bibnamefont
  {{Huelga}}}\ and\ \bibinfo {author} {\bibfnamefont {M.~B.}\ \bibnamefont
  {{Plenio}}},\ }\href@noop {} {\bibfield  {journal} {\bibinfo  {journal}
  {Contemporary Physics}\ }\textbf {\bibinfo {volume} {54}},\ \bibinfo {pages}
  {181} (\bibinfo {year} {2013})}\BibitemShut {NoStop}%
\bibitem [{\citenamefont {{Chin}}\ \emph {et~al.}(2012)\citenamefont {{Chin}},
  \citenamefont {{Huelga}},\ and\ \citenamefont
  {{Plenio}}}]{ChinHuelgaPlenio2012}%
  \BibitemOpen
  \bibfield  {author} {\bibinfo {author} {\bibfnamefont {A.~W.}\ \bibnamefont
  {{Chin}}}, \bibinfo {author} {\bibfnamefont {S.~F.}\ \bibnamefont
  {{Huelga}}}, \ and\ \bibinfo {author} {\bibfnamefont {M.~B.}\ \bibnamefont
  {{Plenio}}},\ }\href@noop {} {\bibfield  {journal} {\bibinfo  {journal}
  {Philosophical Transactions of the Royal Society A: Mathematical, Physical
  and Engineering Sciences}\ }\textbf {\bibinfo {volume} {370}},\ \bibinfo
  {pages} {3638} (\bibinfo {year} {2012})}\BibitemShut {NoStop}%
\bibitem [{\citenamefont {{Womick}}\ and\ \citenamefont
  {{Moran}}(2011)}]{Womick2011}%
  \BibitemOpen
  \bibfield  {author} {\bibinfo {author} {\bibfnamefont {J.~M.}\ \bibnamefont
  {{Womick}}}\ and\ \bibinfo {author} {\bibfnamefont {A.~M.}\ \bibnamefont
  {{Moran}}},\ }\href {\doibase 10.1021/jp106713q} {\bibfield  {journal}
  {\bibinfo  {journal} {The Journal of Physical Chemistry B}\ }\textbf
  {\bibinfo {volume} {115}},\ \bibinfo {pages} {1347} (\bibinfo {year}
  {2011})},\ \Eprint
  {http://arxiv.org/abs/http://pubs.acs.org/doi/pdf/10.1021/jp106713q}
  {http://pubs.acs.org/doi/pdf/10.1021/jp106713q} \BibitemShut {NoStop}%
\bibitem [{\citenamefont {{Hildner}}\ \emph {et~al.}(2013)\citenamefont
  {{Hildner}}, \citenamefont {{Brinks}}, \citenamefont {{Nieder}},
  \citenamefont {{Cogdell}},\ and\ \citenamefont {{Van Hulst}}}]{Hildner2013}%
  \BibitemOpen
  \bibfield  {author} {\bibinfo {author} {\bibfnamefont {R.}~\bibnamefont
  {{Hildner}}}, \bibinfo {author} {\bibfnamefont {D.}~\bibnamefont {{Brinks}}},
  \bibinfo {author} {\bibfnamefont {J.~B.}\ \bibnamefont {{Nieder}}}, \bibinfo
  {author} {\bibfnamefont {R.~J.}\ \bibnamefont {{Cogdell}}}, \ and\ \bibinfo
  {author} {\bibfnamefont {N.~F.}\ \bibnamefont {{Van Hulst}}},\ }\href@noop {}
  {\bibfield  {journal} {\bibinfo  {journal} {Science}\ }\textbf {\bibinfo
  {volume} {340}},\ \bibinfo {pages} {1448} (\bibinfo {year}
  {2013})}\BibitemShut {NoStop}%
\bibitem [{\citenamefont {{Cherezov}}\ \emph {et~al.}(2006)\citenamefont
  {{Cherezov}}, \citenamefont {{Clogston}}, \citenamefont {{Papiz}},\ and\
  \citenamefont {{Caffrey}}}]{PDBfile}%
  \BibitemOpen
  \bibfield  {author} {\bibinfo {author} {\bibfnamefont {V.}~\bibnamefont
  {{Cherezov}}}, \bibinfo {author} {\bibfnamefont {J.}~\bibnamefont
  {{Clogston}}}, \bibinfo {author} {\bibfnamefont {M.~Z.}\ \bibnamefont
  {{Papiz}}}, \ and\ \bibinfo {author} {\bibfnamefont {M.}~\bibnamefont
  {{Caffrey}}},\ }\href@noop {} {\bibfield  {journal} {\bibinfo  {journal}
  {Journal of Molecular Biology}\ }\textbf {\bibinfo {volume} {357}},\ \bibinfo
  {pages} {1605 } (\bibinfo {year} {2006})}\BibitemShut {NoStop}%
\bibitem [{\citenamefont {{Cogdell}}\ \emph {et~al.}(2006)\citenamefont
  {{Cogdell}}, \citenamefont {{Gall}},\ and\ \citenamefont
  {{K\"ohler}}}]{Codgell2006}%
  \BibitemOpen
  \bibfield  {author} {\bibinfo {author} {\bibfnamefont {R.~J.}\ \bibnamefont
  {{Cogdell}}}, \bibinfo {author} {\bibfnamefont {A.}~\bibnamefont {{Gall}}}, \
  and\ \bibinfo {author} {\bibfnamefont {J.}~\bibnamefont {{K\"ohler}}},\
  }\href {\doibase 10.1017/S0033583506004434} {\bibfield  {journal} {\bibinfo
  {journal} {Quarterly Reviews of Biophysics}\ }\textbf {\bibinfo {volume}
  {39}},\ \bibinfo {pages} {227} (\bibinfo {year} {2006})}\BibitemShut
  {NoStop}%
\bibitem [{\citenamefont {{Jimenez}}\ \emph {et~al.}(1997)\citenamefont
  {{Jimenez}}, \citenamefont {{van Mourik}}, \citenamefont {{Young Yu}},\ and\
  \citenamefont {{Flemming}}}]{Jimenez_JPCB1997}%
  \BibitemOpen
  \bibfield  {author} {\bibinfo {author} {\bibfnamefont {R.}~\bibnamefont
  {{Jimenez}}}, \bibinfo {author} {\bibfnamefont {F.}~\bibnamefont {{van
  Mourik}}}, \bibinfo {author} {\bibfnamefont {J.}~\bibnamefont {{Young Yu}}},
  \ and\ \bibinfo {author} {\bibfnamefont {G.~R.}\ \bibnamefont {{Flemming}}},\
  }\href@noop {} {\bibfield  {journal} {\bibinfo  {journal} {J. Phys. Chem. B}\
  }\textbf {\bibinfo {volume} {101}},\ \bibinfo {pages} {7350} (\bibinfo {year}
  {1997})}\BibitemShut {NoStop}%
\bibitem [{\citenamefont {{Yu}}\ \emph {et~al.}(1997)\citenamefont {{Yu}},
  \citenamefont {{Yagasawa}}, \citenamefont {{van Grondelle}},\ and\
  \citenamefont {{Flemming}}}]{vanGrondelle_CPL1997}%
  \BibitemOpen
  \bibfield  {author} {\bibinfo {author} {\bibfnamefont {J.~Y.}\ \bibnamefont
  {{Yu}}}, \bibinfo {author} {\bibfnamefont {Y.}~\bibnamefont {{Yagasawa}}},
  \bibinfo {author} {\bibfnamefont {R.}~\bibnamefont {{van Grondelle}}}, \ and\
  \bibinfo {author} {\bibfnamefont {G.~R.}\ \bibnamefont {{Flemming}}},\
  }\href@noop {} {\bibfield  {journal} {\bibinfo  {journal} {Chem. Phys.
  Lett.}\ }\textbf {\bibinfo {volume} {280}},\ \bibinfo {pages} {404} (\bibinfo
  {year} {1997})}\BibitemShut {NoStop}%
\bibitem [{\citenamefont {{Kr\"uger}}\ \emph {et~al.}(1998)\citenamefont
  {{Kr\"uger}}, \citenamefont {{Scholes}},\ and\ \citenamefont
  {{Fleming}}}]{Krueger1998}%
  \BibitemOpen
  \bibfield  {author} {\bibinfo {author} {\bibfnamefont {B.~P.}\ \bibnamefont
  {{Kr\"uger}}}, \bibinfo {author} {\bibfnamefont {G.~D.}\ \bibnamefont
  {{Scholes}}}, \ and\ \bibinfo {author} {\bibfnamefont {G.~R.}\ \bibnamefont
  {{Fleming}}},\ }\href@noop {} {\bibfield  {journal} {\bibinfo  {journal} {The
  Journal of Physical Chemistry B}\ }\textbf {\bibinfo {volume} {102}},\
  \bibinfo {pages} {5378} (\bibinfo {year} {1998})}\BibitemShut {NoStop}%
\bibitem [{\citenamefont {{Koolhaas}}\ \emph {et~al.}(1998)\citenamefont
  {{Koolhaas}}, \citenamefont {{Frese}}, \citenamefont {{Fowler}},
  \citenamefont {{Bibby}}, \citenamefont {{Georgakopoulou}}, \citenamefont
  {{van der Zwan}}, \citenamefont {{Hunter}},\ and\ \citenamefont
  {R.}}]{vanGrondelle_1998Biochem}%
  \BibitemOpen
  \bibfield  {author} {\bibinfo {author} {\bibfnamefont {M.~H.~C.}\
  \bibnamefont {{Koolhaas}}}, \bibinfo {author} {\bibfnamefont {R.~N.}\
  \bibnamefont {{Frese}}}, \bibinfo {author} {\bibfnamefont {G.~J.~S.}\
  \bibnamefont {{Fowler}}}, \bibinfo {author} {\bibfnamefont {T.~S.}\
  \bibnamefont {{Bibby}}}, \bibinfo {author} {\bibfnamefont {S.}~\bibnamefont
  {{Georgakopoulou}}}, \bibinfo {author} {\bibfnamefont {G.}~\bibnamefont {{van
  der Zwan}}}, \bibinfo {author} {\bibfnamefont {C.~N.}\ \bibnamefont
  {{Hunter}}}, \ and\ \bibinfo {author} {\bibfnamefont {v.}~\bibnamefont
  {R.}},\ }\href@noop {} {\bibfield  {journal} {\bibinfo  {journal}
  {Biochemistry}\ }\textbf {\bibinfo {volume} {37}},\ \bibinfo {pages} {4693}
  (\bibinfo {year} {1998})}\BibitemShut {NoStop}%
\bibitem [{\citenamefont {{Sundstr\"om}}\ \emph {et~al.}(1999)\citenamefont
  {{Sundstr\"om}}, \citenamefont {{Pullerits}},\ and\ \citenamefont {{Van
  Grondelle}}}]{Sundstrom}%
  \BibitemOpen
  \bibfield  {author} {\bibinfo {author} {\bibfnamefont {V.}~\bibnamefont
  {{Sundstr\"om}}}, \bibinfo {author} {\bibfnamefont {T.}~\bibnamefont
  {{Pullerits}}}, \ and\ \bibinfo {author} {\bibfnamefont {R.}~\bibnamefont
  {{Van Grondelle}}},\ }\href@noop {} {\bibfield  {journal} {\bibinfo
  {journal} {The Journal of Physical Chemistry B}\ }\textbf {\bibinfo {volume}
  {103}},\ \bibinfo {pages} {2327} (\bibinfo {year} {1999})}\BibitemShut
  {NoStop}%
\bibitem [{\citenamefont {{van Oijen}}\ \emph {et~al.}(2000)\citenamefont {{van
  Oijen}}, \citenamefont {{Katelaars}}, \citenamefont {{K\"oler}},
  \citenamefont {{Aartsma}},\ and\ \citenamefont
  {{Schmidt}}}]{vanOIjeen_BioPhys2000}%
  \BibitemOpen
  \bibfield  {author} {\bibinfo {author} {\bibfnamefont {A.~M.}\ \bibnamefont
  {{van Oijen}}}, \bibinfo {author} {\bibfnamefont {M.}~\bibnamefont
  {{Katelaars}}}, \bibinfo {author} {\bibfnamefont {J.}~\bibnamefont
  {{K\"oler}}}, \bibinfo {author} {\bibfnamefont {T.~J.}\ \bibnamefont
  {{Aartsma}}}, \ and\ \bibinfo {author} {\bibfnamefont {J.}~\bibnamefont
  {{Schmidt}}},\ }\href@noop {} {\bibfield  {journal} {\bibinfo  {journal}
  {Biophys. J.}\ }\textbf {\bibinfo {volume} {78}},\ \bibinfo {pages} {1570}
  (\bibinfo {year} {2000})}\BibitemShut {NoStop}%
\bibitem [{\citenamefont {{Monshouwer}}\ \emph {et~al.}(1997)\citenamefont
  {{Monshouwer}}, \citenamefont {{Abrahamsson}}, \citenamefont {{van Mourik}},\
  and\ \citenamefont {{van Grondelle}}}]{vanGrondelle_1997JPCB}%
  \BibitemOpen
  \bibfield  {author} {\bibinfo {author} {\bibfnamefont {R.}~\bibnamefont
  {{Monshouwer}}}, \bibinfo {author} {\bibfnamefont {M.}~\bibnamefont
  {{Abrahamsson}}}, \bibinfo {author} {\bibfnamefont {F.}~\bibnamefont {{van
  Mourik}}}, \ and\ \bibinfo {author} {\bibfnamefont {R.}~\bibnamefont {{van
  Grondelle}}},\ }\href@noop {} {\bibfield  {journal} {\bibinfo  {journal} {J.
  Phys. Chem. B}\ }\textbf {\bibinfo {volume} {101}},\ \bibinfo {pages} {7241}
  (\bibinfo {year} {1997})}\BibitemShut {NoStop}%
\bibitem [{\citenamefont {{Hu}}\ \emph {et~al.}(2002)\citenamefont {{Hu}},
  \citenamefont {{Ritz}}, \citenamefont {{Damjanovi\'c}}, \citenamefont
  {{Authennrieth}},\ and\ \citenamefont {{Schulten}}}]{Hu_2002}%
  \BibitemOpen
  \bibfield  {author} {\bibinfo {author} {\bibfnamefont {X.}~\bibnamefont
  {{Hu}}}, \bibinfo {author} {\bibfnamefont {T.}~\bibnamefont {{Ritz}}},
  \bibinfo {author} {\bibfnamefont {A.}~\bibnamefont {{Damjanovi\'c}}},
  \bibinfo {author} {\bibfnamefont {F.}~\bibnamefont {{Authennrieth}}}, \ and\
  \bibinfo {author} {\bibfnamefont {K.}~\bibnamefont {{Schulten}}},\
  }\href@noop {} {\bibfield  {journal} {\bibinfo  {journal} {Quarterly Review
  of Biophysics}\ }\textbf {\bibinfo {volume} {35}},\ \bibinfo {pages} {1}
  (\bibinfo {year} {2002})}\BibitemShut {NoStop}%
\bibitem [{\citenamefont {{Trinkunas}}\ and\ \citenamefont
  {{Freiberg}}(2006)}]{Trinkunas_JLum_2003}%
  \BibitemOpen
  \bibfield  {author} {\bibinfo {author} {\bibfnamefont {G.}~\bibnamefont
  {{Trinkunas}}}\ and\ \bibinfo {author} {\bibfnamefont {A.}~\bibnamefont
  {{Freiberg}}},\ }\href@noop {} {\bibfield  {journal} {\bibinfo  {journal}
  {Journal of Luminescence}\ }\textbf {\bibinfo {volume} {119-120}},\ \bibinfo
  {pages} {105} (\bibinfo {year} {2006})}\BibitemShut {NoStop}%
\bibitem [{\citenamefont {{Timpmann}}\ \emph {et~al.}(2004)\citenamefont
  {{Timpmann}}, \citenamefont {{Trinkunas}}, \citenamefont {{Olsen}},
  \citenamefont {{Hunter}},\ and\ \citenamefont {{Freiberg}}}]{Timpmann2004}%
  \BibitemOpen
  \bibfield  {author} {\bibinfo {author} {\bibfnamefont {K.}~\bibnamefont
  {{Timpmann}}}, \bibinfo {author} {\bibfnamefont {G.}~\bibnamefont
  {{Trinkunas}}}, \bibinfo {author} {\bibfnamefont {J.~D.}\ \bibnamefont
  {{Olsen}}}, \bibinfo {author} {\bibfnamefont {C.~N.}\ \bibnamefont
  {{Hunter}}}, \ and\ \bibinfo {author} {\bibfnamefont {A.}~\bibnamefont
  {{Freiberg}}},\ }\href@noop {} {\bibfield  {journal} {\bibinfo  {journal}
  {Chemical Physics Letters}\ }\textbf {\bibinfo {volume} {398}},\ \bibinfo
  {pages} {384 } (\bibinfo {year} {2004})}\BibitemShut {NoStop}%
\bibitem [{\citenamefont {Urboniene}\ \emph {et~al.}(2005)\citenamefont
  {Urboniene}, \citenamefont {Vrublevskaja}, \citenamefont {Gall},
  \citenamefont {Trinkunas}, \citenamefont {Robert},\ and\ \citenamefont
  {Valkunas}}]{Urboniene_2005Pres}%
  \BibitemOpen
  \bibfield  {author} {\bibinfo {author} {\bibfnamefont {V.}~\bibnamefont
  {Urboniene}}, \bibinfo {author} {\bibfnamefont {O.}~\bibnamefont
  {Vrublevskaja}}, \bibinfo {author} {\bibfnamefont {A.}~\bibnamefont {Gall}},
  \bibinfo {author} {\bibfnamefont {G.}~\bibnamefont {Trinkunas}}, \bibinfo
  {author} {\bibfnamefont {B.}~\bibnamefont {Robert}}, \ and\ \bibinfo {author}
  {\bibfnamefont {L.}~\bibnamefont {Valkunas}},\ }\href {\doibase
  10.1007/s11120-005-2748-9} {\bibfield  {journal} {\bibinfo  {journal}
  {Photosynthesis Research}\ }\textbf {\bibinfo {volume} {86}},\ \bibinfo
  {pages} {49} (\bibinfo {year} {2005})}\BibitemShut {NoStop}%
\bibitem [{\citenamefont {{Cheng}}\ and\ \citenamefont
  {{Silbey}}(2006)}]{Silbey_PRL_2006}%
  \BibitemOpen
  \bibfield  {author} {\bibinfo {author} {\bibfnamefont {R.~C.}\ \bibnamefont
  {{Cheng}}}\ and\ \bibinfo {author} {\bibfnamefont {R.~J.}\ \bibnamefont
  {{Silbey}}},\ }\href@noop {} {\bibfield  {journal} {\bibinfo  {journal}
  {Phys. Rev. Lett.}\ }\textbf {\bibinfo {volume} {96}},\ \bibinfo {pages}
  {028103} (\bibinfo {year} {2006})}\BibitemShut {NoStop}%
\bibitem [{\citenamefont {{Mal\'y}}\ \emph {et~al.}(2016)\citenamefont
  {{Mal\'y}}, \citenamefont {{Gruber}}, \citenamefont {{Cogdell}},
  \citenamefont {{Man{\v c}al}},\ and\ \citenamefont {{van
  Grondelle}}}]{vanGrondelle_PNAS2016}%
  \BibitemOpen
  \bibfield  {author} {\bibinfo {author} {\bibfnamefont {P.}~\bibnamefont
  {{Mal\'y}}}, \bibinfo {author} {\bibfnamefont {J.~M.}\ \bibnamefont
  {{Gruber}}}, \bibinfo {author} {\bibfnamefont {R.~J.}\ \bibnamefont
  {{Cogdell}}}, \bibinfo {author} {\bibfnamefont {T.}~\bibnamefont {{Man{\v
  c}al}}}, \ and\ \bibinfo {author} {\bibnamefont {{van Grondelle}}},\
  }\href@noop {} {\bibfield  {journal} {\bibinfo  {journal} {PNAS}\ }\textbf
  {\bibinfo {volume} {113}},\ \bibinfo {pages} {2934} (\bibinfo {year}
  {2016})}\BibitemShut {NoStop}%
\bibitem [{\citenamefont {{Pullerits}}\ \emph {et~al.}(1997)\citenamefont
  {{Pullerits}}, \citenamefont {{Hess}}, \citenamefont {{Herek}},\ and\
  \citenamefont {{Sundstr\"om}}}]{Pullerits1997}%
  \BibitemOpen
  \bibfield  {author} {\bibinfo {author} {\bibfnamefont {T.}~\bibnamefont
  {{Pullerits}}}, \bibinfo {author} {\bibfnamefont {S.}~\bibnamefont {{Hess}}},
  \bibinfo {author} {\bibfnamefont {J.~L.}\ \bibnamefont {{Herek}}}, \ and\
  \bibinfo {author} {\bibfnamefont {V.}~\bibnamefont {{Sundstr\"om}}},\
  }\href@noop {} {\bibfield  {journal} {\bibinfo  {journal} {The Journal of
  Physical Chemistry B}\ }\textbf {\bibinfo {volume} {101}},\ \bibinfo {pages}
  {10560} (\bibinfo {year} {1997})}\BibitemShut {NoStop}%
\bibitem [{\citenamefont {{Olbrich}}\ \emph {et~al.}(2011)\citenamefont
  {{Olbrich}}, \citenamefont {{Str\"umpire}}, \citenamefont {{Schulten}},\ and\
  \citenamefont {{Kleinekath\"ofer}}}]{Olbrich2011}%
  \BibitemOpen
  \bibfield  {author} {\bibinfo {author} {\bibfnamefont {C.}~\bibnamefont
  {{Olbrich}}}, \bibinfo {author} {\bibfnamefont {J.}~\bibnamefont
  {{Str\"umpire}}}, \bibinfo {author} {\bibfnamefont {K.}~\bibnamefont
  {{Schulten}}}, \ and\ \bibinfo {author} {\bibfnamefont {U.}~\bibnamefont
  {{Kleinekath\"ofer}}},\ }\href@noop {} {\bibfield  {journal} {\bibinfo
  {journal} {J. Phys. Chem. B}\ }\textbf {\bibinfo {volume} {115}},\ \bibinfo
  {pages} {758} (\bibinfo {year} {2011})}\BibitemShut {NoStop}%
\bibitem [{\citenamefont {{Van Amerongen}}\ \emph {et~al.}(2000)\citenamefont
  {{Van Amerongen}}, \citenamefont {{Valkaunas}},\ and\ \citenamefont {{Van
  Grondelle}}}]{VanGrondelleExcitonsBook}%
  \BibitemOpen
  \bibfield  {author} {\bibinfo {author} {\bibfnamefont {H.}~\bibnamefont {{Van
  Amerongen}}}, \bibinfo {author} {\bibfnamefont {L.}~\bibnamefont
  {{Valkaunas}}}, \ and\ \bibinfo {author} {\bibfnamefont {R.}~\bibnamefont
  {{Van Grondelle}}},\ }\href@noop {} {\emph {\bibinfo {title} {{Photosynthetic
  Excitons}}}}\ (\bibinfo  {publisher} {World Scientific},\ \bibinfo {year}
  {2000})\BibitemShut {NoStop}%
\bibitem [{\citenamefont {{Lim}}\ \emph {et~al.}(2015)\citenamefont {{Lim}},
  \citenamefont {Pale{\v c}ek}, \citenamefont {Caycedo-Soler}, \citenamefont
  {Lincoln}, \citenamefont {Prior}, \citenamefont {von Berlepsch},
  \citenamefont {Huelga}, \citenamefont {Plenio}, \citenamefont {Zigmantas},\
  and\ \citenamefont {Hauer}}]{Lim_2015NC}%
  \BibitemOpen
  \bibfield  {author} {\bibinfo {author} {\bibfnamefont {J.}~\bibnamefont
  {{Lim}}}, \bibinfo {author} {\bibfnamefont {D.}~\bibnamefont {Pale{\v c}ek}},
  \bibinfo {author} {\bibfnamefont {F.}~\bibnamefont {Caycedo-Soler}}, \bibinfo
  {author} {\bibfnamefont {C.~N.}\ \bibnamefont {Lincoln}}, \bibinfo {author}
  {\bibfnamefont {J.}~\bibnamefont {Prior}}, \bibinfo {author} {\bibfnamefont
  {H.}~\bibnamefont {von Berlepsch}}, \bibinfo {author} {\bibfnamefont {S.~F.}\
  \bibnamefont {Huelga}}, \bibinfo {author} {\bibfnamefont {M.~B.}\
  \bibnamefont {Plenio}}, \bibinfo {author} {\bibfnamefont {D.}~\bibnamefont
  {Zigmantas}}, \ and\ \bibinfo {author} {\bibfnamefont {J.}~\bibnamefont
  {Hauer}},\ }\href@noop {} {\bibfield  {journal} {\bibinfo  {journal}
  {arXiv:1502.01717}\ } (\bibinfo {year} {2015})}\BibitemShut {NoStop}%
\bibitem [{\citenamefont {{Breuer}}\ and\ \citenamefont
  {{Petrucione}}(2002)}]{BreuerPetrucione}%
  \BibitemOpen
  \bibfield  {author} {\bibinfo {author} {\bibfnamefont {H.~P.}\ \bibnamefont
  {{Breuer}}}\ and\ \bibinfo {author} {\bibfnamefont {F.}~\bibnamefont
  {{Petrucione}}},\ }\href@noop {} {\emph {\bibinfo {title} {{The theory of
  open quantum systems}}}}\ (\bibinfo  {publisher} {Oxford},\ \bibinfo {year}
  {2002})\BibitemShut {NoStop}%
\bibitem [{\citenamefont {{van Amerongen}}\ \emph {et~al.}(2000)\citenamefont
  {{van Amerongen}}, \citenamefont {{Valkunas}},\ and\ \citenamefont {{van
  Grondelle}}}]{PhotoEx}%
  \BibitemOpen
  \bibfield  {author} {\bibinfo {author} {\bibfnamefont {H.}~\bibnamefont {{van
  Amerongen}}}, \bibinfo {author} {\bibfnamefont {L.}~\bibnamefont
  {{Valkunas}}}, \ and\ \bibinfo {author} {\bibfnamefont {R.}~\bibnamefont
  {{van Grondelle}}},\ }\href@noop {} {\emph {\bibinfo {title} {{Photosynthetic
  Excitons}}}}\ (\bibinfo  {publisher} {World Scientific Publishing Co. Pte.
  Lrd.},\ \bibinfo {year} {2000})\BibitemShut {NoStop}%
\bibitem [{\citenamefont {{Reddy}}\ \emph {et~al.}(1991)\citenamefont
  {{Reddy}}, \citenamefont {{Small}}, \citenamefont {{Seibert}},\ and\
  \citenamefont {{Picorel}}}]{Reddy1991}%
  \BibitemOpen
  \bibfield  {author} {\bibinfo {author} {\bibfnamefont {N.}~\bibnamefont
  {{Reddy}}}, \bibinfo {author} {\bibfnamefont {G.}~\bibnamefont {{Small}}},
  \bibinfo {author} {\bibfnamefont {M.}~\bibnamefont {{Seibert}}}, \ and\
  \bibinfo {author} {\bibfnamefont {R.}~\bibnamefont {{Picorel}}},\ }\href@noop
  {} {\bibfield  {journal} {\bibinfo  {journal} {Chemical Physics Letters}\
  }\textbf {\bibinfo {volume} {181}},\ \bibinfo {pages} {391 } (\bibinfo {year}
  {1991})}\BibitemShut {NoStop}%
\bibitem [{\citenamefont {I.~{Renge}}\ \emph {et~al.}(1987)\citenamefont
  {I.~{Renge}}, \citenamefont {{Mauring}},\ and\ \citenamefont
  {{Avarmaa}}}]{Renge_1987JLum}%
  \BibitemOpen
  \bibfield  {author} {\bibinfo {author} {\bibfnamefont {I.}~\bibnamefont
  {I.~{Renge}}}, \bibinfo {author} {\bibfnamefont {K.}~\bibnamefont
  {{Mauring}}}, \ and\ \bibinfo {author} {\bibfnamefont {R.}~\bibnamefont
  {{Avarmaa}}},\ }\href {\doibase https://doi.org/10.1016/0022-2313(87)90161-X}
  {\bibfield  {journal} {\bibinfo  {journal} {Journal of Luminescence}\
  }\textbf {\bibinfo {volume} {37}},\ \bibinfo {pages} {207 } (\bibinfo {year}
  {1987})}\BibitemShut {NoStop}%
\bibitem [{\citenamefont {{Mukamel}}(1995)}]{Mukamel}%
  \BibitemOpen
  \bibfield  {author} {\bibinfo {author} {\bibfnamefont {S.}~\bibnamefont
  {{Mukamel}}},\ }\href@noop {} {\emph {\bibinfo {title} {Principles of
  Nonlinear Optical Spectroscopy}}}\ (\bibinfo  {publisher} {Oxford},\ \bibinfo
  {year} {1995})\BibitemShut {NoStop}%
\bibitem [{\citenamefont {{Caycedo-Soler}}\ \emph {et~al.}(2017)\citenamefont
  {{Caycedo-Soler}}, \citenamefont {{Schroeder}}, \citenamefont {{Autenrieth}},
  \citenamefont {{Pick}}, \citenamefont {{Ghosh}}, \citenamefont {{Huelga}},\
  and\ \citenamefont {{Plenio}}}]{Caycedo_JPC2017}%
  \BibitemOpen
  \bibfield  {author} {\bibinfo {author} {\bibfnamefont {F.}~\bibnamefont
  {{Caycedo-Soler}}}, \bibinfo {author} {\bibfnamefont {C.~A.}\ \bibnamefont
  {{Schroeder}}}, \bibinfo {author} {\bibfnamefont {C.}~\bibnamefont
  {{Autenrieth}}}, \bibinfo {author} {\bibfnamefont {A.}~\bibnamefont
  {{Pick}}}, \bibinfo {author} {\bibfnamefont {R.}~\bibnamefont {{Ghosh}}},
  \bibinfo {author} {\bibfnamefont {S.~F.}\ \bibnamefont {{Huelga}}}, \ and\
  \bibinfo {author} {\bibfnamefont {M.~B.}\ \bibnamefont {{Plenio}}},\
  }\href@noop {} {\bibfield  {journal} {\bibinfo  {journal} {J. Phys. Chem.
  Lett.}\ }\textbf {\bibinfo {volume} {8}},\ \bibinfo {pages} {6015} (\bibinfo
  {year} {2017})}\BibitemShut {NoStop}%
\bibitem [{\citenamefont {{Cohen-Stuart}}\ \emph {et~al.}(2011)\citenamefont
  {{Cohen-Stuart}}, \citenamefont {{Vengris}}, \citenamefont {{Novoderezhkin}},
  \citenamefont {{Cogdell}}, \citenamefont {{Hunter}},\ and\ \citenamefont
  {{van Grondelle}}}]{Cohen_2011BioJ}%
  \BibitemOpen
  \bibfield  {author} {\bibinfo {author} {\bibfnamefont {T.~A.}\ \bibnamefont
  {{Cohen-Stuart}}}, \bibinfo {author} {\bibfnamefont {M.}~\bibnamefont
  {{Vengris}}}, \bibinfo {author} {\bibfnamefont {V.~I.}\ \bibnamefont
  {{Novoderezhkin}}}, \bibinfo {author} {\bibfnamefont {R.~J.}\ \bibnamefont
  {{Cogdell}}}, \bibinfo {author} {\bibfnamefont {C.~N.}\ \bibnamefont
  {{Hunter}}}, \ and\ \bibinfo {author} {\bibfnamefont {R.}~\bibnamefont {{van
  Grondelle}}},\ }\href {\doibase 10.1016/j.bpj.2011.02.048} {\bibfield
  {journal} {\bibinfo  {journal} {Biophysical Journal}\ }\textbf {\bibinfo
  {volume} {100}},\ \bibinfo {pages} {2226 } (\bibinfo {year}
  {2011})}\BibitemShut {NoStop}%
\bibitem [{Cum()}]{Cumulant}%
  \BibitemOpen
  \href@noop {} {\ }\bibinfo {note} {The cumulant expansion can not --in
  principle-- be used if the exciton coupling is similar to the e-ph coupling.
  However, in the limiting case of strong e-ph coupling, as compared to the
  inverse of the bath correlation time, Gaussians can be used to obtain
  expressions for the lineshapes thus fitting the spectrum}\BibitemShut
  {NoStop}%
\bibitem [{\citenamefont {{Katelaars}}\ \emph {et~al.}(2001)\citenamefont
  {{Katelaars}}, \citenamefont {{van Oijen}}, \citenamefont {{Matsushita}},
  \citenamefont {{K\"oler}}, \citenamefont {{Schmidt}},\ and\ \citenamefont
  {{Aartsma}}}]{Katelaars_Biophys2001}%
  \BibitemOpen
  \bibfield  {author} {\bibinfo {author} {\bibfnamefont {M.}~\bibnamefont
  {{Katelaars}}}, \bibinfo {author} {\bibfnamefont {A.~M.}\ \bibnamefont {{van
  Oijen}}}, \bibinfo {author} {\bibfnamefont {M.}~\bibnamefont {{Matsushita}}},
  \bibinfo {author} {\bibfnamefont {J.}~\bibnamefont {{K\"oler}}}, \bibinfo
  {author} {\bibfnamefont {J.}~\bibnamefont {{Schmidt}}}, \ and\ \bibinfo
  {author} {\bibfnamefont {T.~J.}\ \bibnamefont {{Aartsma}}},\ }\href@noop {}
  {\bibfield  {journal} {\bibinfo  {journal} {Biophys. J.}\ }\textbf {\bibinfo
  {volume} {80}},\ \bibinfo {pages} {1591} (\bibinfo {year}
  {2001})}\BibitemShut {NoStop}%
\bibitem [{\citenamefont {Rivas}\ and\ \citenamefont
  {Huelga}(2011)}]{RivasHuelga_2010}%
  \BibitemOpen
  \bibfield  {author} {\bibinfo {author} {\bibfnamefont {{\'A}.}~\bibnamefont
  {Rivas}}\ and\ \bibinfo {author} {\bibfnamefont {S.~F.}\ \bibnamefont
  {Huelga}},\ }\href@noop {} {\emph {\bibinfo {title} {Open Quantum Systems. An
  Introduction}}}\ (\bibinfo  {publisher} {Heidelberg: Springer},\ \bibinfo
  {year} {2011})\BibitemShut {NoStop}%
\bibitem [{\citenamefont {Tsomokos}\ \emph {et~al.}(2007)\citenamefont
  {Tsomokos}, \citenamefont {Hartmann}, \citenamefont {Huelga},\ and\
  \citenamefont {Plenio}}]{1367-2630-9-3-079}%
  \BibitemOpen
  \bibfield  {author} {\bibinfo {author} {\bibfnamefont {D.~I.}\ \bibnamefont
  {Tsomokos}}, \bibinfo {author} {\bibfnamefont {M.~J.}\ \bibnamefont
  {Hartmann}}, \bibinfo {author} {\bibfnamefont {S.~F.}\ \bibnamefont
  {Huelga}}, \ and\ \bibinfo {author} {\bibfnamefont {M.~B.}\ \bibnamefont
  {Plenio}},\ }\href {http://stacks.iop.org/1367-2630/9/i=3/a=079} {\bibfield
  {journal} {\bibinfo  {journal} {New Journal of Physics}\ }\textbf {\bibinfo
  {volume} {9}},\ \bibinfo {pages} {79} (\bibinfo {year} {2007})}\BibitemShut
  {NoStop}%
\end{thebibliography}%

\end{document}